\documentclass[onecolumn, preprintnumbers,amsmath,amssymb, preprint]{revtex4}

\usepackage{graphicx}
\usepackage{dcolumn}
\usepackage{bm}

\usepackage[usenames]{color}

\newcommand\blbr[1]{{\color{black} #1}}

\begin{document}
\preprint{}

\title{Helical Swimming in Stokes Flow Using a Novel Boundary-element Method}

\author{Bin Liu$^{1}$, Kenneth~S. Breuer$^{1}$, Thomas~R.~Powers$^{1,2}$}
\affiliation{$^1$School of Engineering}
\affiliation{$^2$Department of Physics, Brown University, Providence, Rhode Island 02912, USA}

\date{\today}
\begin{abstract}
We apply the boundary-element method to Stokes flows 
with helical symmetry, 
such as the flow driven by an immersed rotating helical flagellum. 
We show that the 
two-dimensional boundary integral method can be reduced to one dimension using the helical symmetry. 
The computational cost is thus much reduced while spatial resolution is maintained. We review the robustness of this method by comparing the simulation results with the experimental measurement of the motility of model helical flagella of various ratios of pitch to radius, along with predictions from resistive-force theory and slender-body theory. We also show that the modified boundary integral method provides reliable convergence if the singularities in the kernel of the integral are treated appropriately.

\end{abstract}



\maketitle

\section{Introduction}

The boundary-element method (BEM)~\cite{Pozrikidis:1992td} is 
a reliable and accurate tool for 
studying the zero-Reynolds number hydrodynamics of motile microorganisms~\cite{Lauga:RPP2009}, especially in situations where the detailed geometry of the structure of the microorganism plays a role~\cite{Phan-Thien:1987}. When properly formulated, the 
BEM leads to robust solutions with high orders of convergence~\cite{Power:1987, Gonzalez:2009, Keaveny:JCP2011}. However, the trade-off for high reliability is that the computational cost of this method is typically high, increasing drastically with the number of mesh elements used for spatial discretization. In practice, this heavy computational load can be reduced either by approximating the singular kernel in a regularized form, as in the method of regularized Stokeslets~\cite{Cortez:2001, Cortez:PF2005} or by replacing the force density with a piecewise constant function~\cite{Smith:2009kh}. With these approximations, fewer nodes are required for a given boundary geometry. However, to achieve robust convergence, regularizing the singular kernel requires a delicate selection of the numerical parameters describing the width and spacing of the regularized Stokeslets~\cite{Cortez:PF2005, Bouzarth:2011}.
Likewise, there is no natural prescription for assigning the patches of constant force, and clever choices of the constant-force patches are often necessary~\cite{Smith:2009kh}. 
However, small motile organisms often have symmetric bodies or ciliary beat patterns, and these symmetries have not always been exploited in computation. 
For instance, bacteria such as \textit{Eshericia coli} swim by rotating helical flagella \cite{Berg:1973td}, and the spirochete \textit{Leptonema illini} has a helically shaped cell body \cite{Charon:1984}. These structures typically have many pitches, and thus approximate perfect helical symmetry. 
Another example is the array of beating cilia on an actively swimming \textit{Paramecium}, which coordinate to form metachronal wave pattens. This wavefront follows a counterclockwise gyration along the cell body \cite{Horridge:1969vt}, again leading to an approximate helical symmetry. 

Symmetries in fluid-structure interaction problems 
have been used by others to reduce the number of unknowns and thus simplify numerical computation. For an axially symmetric low-Reynolds-number swimmer~\cite{Purcell:1977}, the Green's function in a boundary-integral method is modified to include integration along the azimuthal direction~\cite{Pozrikidis:1992td}, and the two-dimensional (2$d$) boundary-integral problem can thus be reduced to a one-dimension (1$d$) problem~\cite{Thaokar:2007, Spagnolie:2010}. Helical symmetry has been applied in Lighthill's slender-body-theory calculation of the motility of rotating flagella \cite{Lighthill:1976,lighthill1996b}, where every point on the helical body-centerline is regarded as identical. More generally, it has been shown that the 3$d$ Navier-Stokes equation for flow with helical symmetry can be reduced to a 2$d$ problem \cite{Childress:1989}. 
This technique was used to study flow within a helical pipe \cite{Childress:1989, Zabielski:1998}, and the helical wake of a rotating propeller at large Reynolds numbers \cite{Delbende:2011}. 

In this study we show, in general, that symmetry in the surface domain leads to a reduction in dimension of  the boundary-integral method. More specifically, for a helical symmetry, a 2$d$ boundary integral can be reduced to 1$d$. The idea of this dimension reduction is straightforward: if we know the force vectors on an arbitrary circumference around a helical body, we know immediately the force distribution on the entire surface of the body, since the whole surface can be reconstructed from these identical circumferences. 
The computational benefit from exploiting this symmetry is evident.
Suppose the surface of the helix is approximated by a mesh with $N_\alpha$ nodes around  each circumference, and $N_\varphi$ circumferences, for a total number of mesh points of $N_\alpha N_\varphi$. In a full boundary-element method, we are required to invert a $3N_\alpha N_\varphi\times3N_\alpha N_\varphi$ matrix. By exploiting the helical symmetry we reduce the problem to that of inverting a much smaller $3N_\alpha\times3N_\alpha$ matrix.

As an example application, we use this modified boundary-element method to model the motion of a bacterium propelled by a helical flagellum, and compute the swimming speed of a rotating helix subject to zero force. We demonstrate that our
numerical method is highly reliable, and achieves third-order convergence with proper singularity reduction. We also compare quantitatively our simulation results with some experimental studies on the motility of model helical swimmers, along with those predicted from other theoretical tools, such as  resistive force theory and slender-body theory. 

\section{Helical Symmetry and the Boundary Integral Method}

\begin{figure}
\begin{center}
\includegraphics[width=0.8\textwidth]{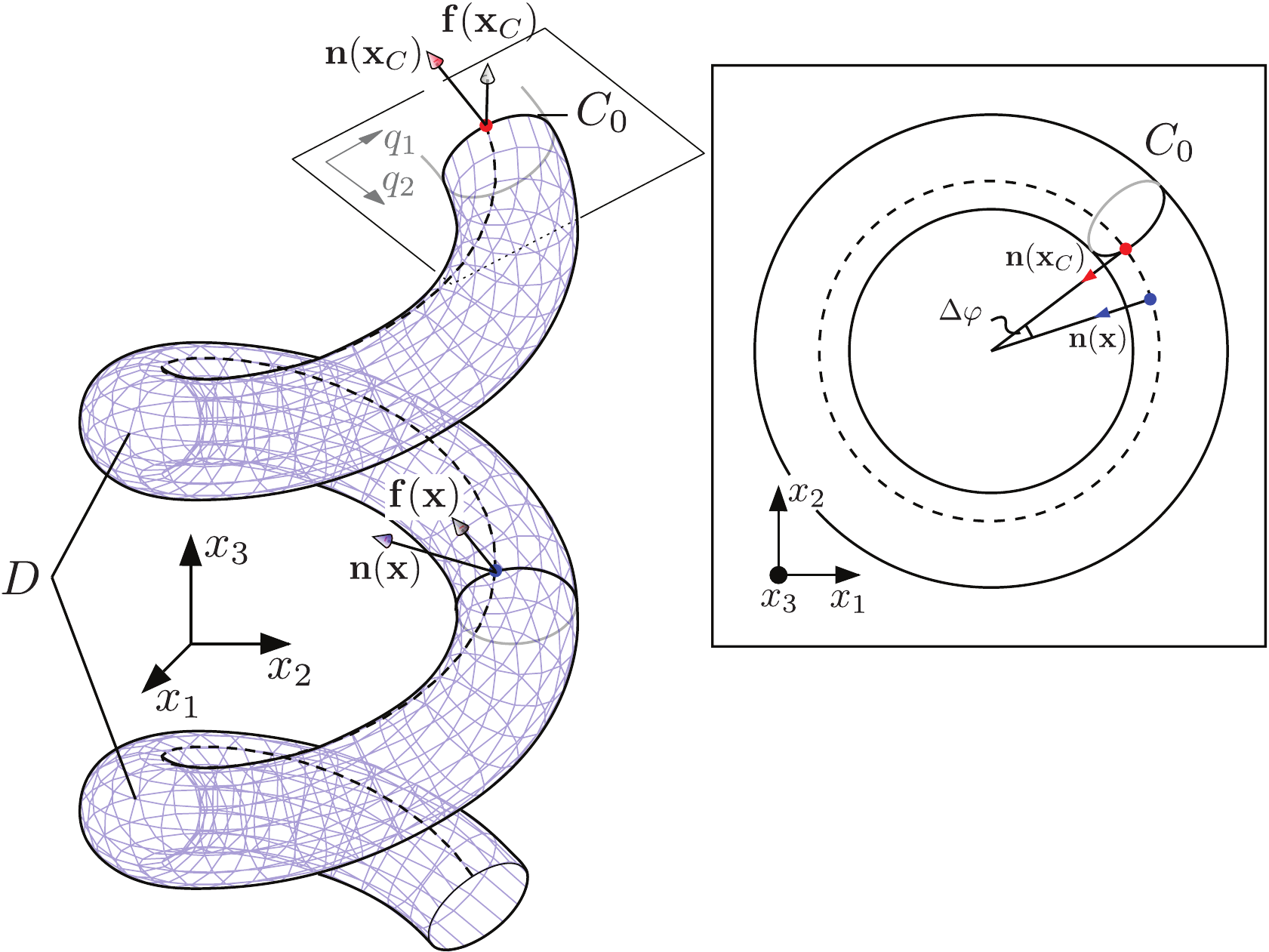}
\caption{(Color online.) The value $\mathbf{f}(\mathbf{x})$ of a vector field at an arbitrary location on a 
surface $D$  with helical symmetry can be determined 
from the vector $\mathbf{f}(\mathbf{x}_C)$  
on a given circumference $C_0$ through a rotation about the axis $x_3$. The rotation angle $\Delta \varphi$ is 
the angle 
between the projections of the surface normals at positions $\mathbf{x}$ and $\mathbf{x}_C$ in the $x_1$--$x_2$ plane. The dashed curve shows a contour along which vector fields $\mathbf{f}(\mathbf{x})$ have the 
same magnitude but vary in orientation. The inset shows a view of the helical filament along its axis of
symmetry. 
\label{fig:1}}
\end{center}
\end{figure}

In the boundary-integral method, the flow field $\mathbf{u}$  of a Stokes fluid can be represented in terms of two boundary distributions \cite{Pozrikidis:1992td}, involving the Green's function $\mathcal{G}$, the associated stresslet $\mathcal{T}$, and the force density $\mathbf{f}$ 
on the domain  $D$,  
\begin{equation}
\begin{array}{ll}
u_j(\mathbf{x})=&-\frac{1}{8\pi\mu}\int_D \textrm{d}S_{\mathbf{x'}} f_i(\mathbf{x}')\mathcal{G}_{ij}(\mathbf{x}', \mathbf{x}) \\
&+\frac{1}{8\pi}\int_D \textrm{d}S_{\mathbf{x'}} u_i(\mathbf{x'}) T_{ijk} (\mathbf{x}', \mathbf{x})n_k(\mathbf{x}').
\end{array} \label{eq:bi}
\end{equation}
The first distribution on the right-hand side of Eq.~(\ref{eq:bi}) is 
the single-layer potential, 
and the second distribution is the double-layer potential. The Green's function $\mathcal{G}$, 
also known as the Stokeslet, is 
\begin{equation}
\mathcal{G}_{ij}(\mathbf{x}', \mathbf{x})=\frac{\delta_{ij}}{\left|\mathbf{x}- \mathbf{x}'\right|}+\frac{(x'_i-x_i)(x'_j-x_j)}{\left|\mathbf{x}- \mathbf{x}'\right|^3},\label{greens} 
\end{equation}
and the stresslet is
\begin{equation}
\mathcal{T}_{ijk}(\mathbf{x}', \mathbf{x})=\frac{-6(x'_i-x_i)(x'_j-x_j)(x'_k-x_k)}{\left|\mathbf{x}- \mathbf{x}'\right|^5}. 
\end{equation}

We are interested in the motion of a rigid particle, in which case the velocity may be written solely in terms of the single-layer potential~\cite{kimbook}:
\begin{equation}
u_j(\mathbf{x})=-\frac{1}{8\pi\mu}\int_D \textrm{d}S_{\mathbf{x'}} f_i(\mathbf{x}')\mathcal{G}_{ij}(\mathbf{x}', \mathbf{x}). \label{bi-cartesian}
\end{equation}
The velocity field is linear in the force density, $f_i$. For a prescribed velocity field on the meshed surface, $\mathbf{x}^i\in D$ ($i=1$, 2, $\cdots$, N), the force density, $\mathbf{f}(\mathbf{x}^i)$, can be determined using the discretized version of (Eq. \ref{bi-cartesian}): 

\begin{equation}
\textrm{$\scriptsize
\left(
\begin{array}{c}
\mathbf{u}\left(\mathbf{x}^1\right)\\
\mathbf{u}\left(\mathbf{x}^2\right)\\
\cdots\\
\mathbf{u}\left(\mathbf{x}^N\right)\\
\end{array}
\right)$=$-\frac{1}{8\pi\mu}
\mathcal{L}
\cdot$$\scriptsize
\left(
\begin{array}{c}
\mathbf{f}\left(\mathbf{x}^1\right)\\
\mathbf{f}\left(\mathbf{x}^2\right)\\
\cdots\\
\mathbf{f}\left(\mathbf{x}^N\right)\\
\end{array}
\right)
$},
\end{equation}
where the 
matrix $\mathcal{L}$ is the $3N\times3N$ matrix 
\begin{equation}\textrm{
$\mathcal{L}=$
$\scriptsize
\left(
\begin{array}{cccc}
\mathcal{G} (\mathbf{x}^{1}, \mathbf{x}^{1}) \Delta S(\mathbf{x}^{1}) &\mathcal{G} (\mathbf{x}^{1}, \mathbf{x}^{2}) \Delta S(\mathbf{x}^{2}) &\cdots &\mathcal{G} (\mathbf{x}^{1}, \mathbf{x}^{N}) \Delta S(\mathbf{x}^{N})\\
\mathcal{G} (\mathbf{x}^{2}, \mathbf{x}^{1}) \Delta S(\mathbf{x}^{1}) &\mathcal{G} (\mathbf{x}^{2}, \mathbf{x}^{2}) \Delta S(\mathbf{x}^{2}) &\cdots &\mathcal{G} (\mathbf{x}^{2}, \mathbf{x}^{N}) \Delta S(\mathbf{x}^{N})\\
\cdots &\cdots &\cdots &\cdots\\
\mathcal{G} (\mathbf{x}^{N}, \mathbf{x}^{1}) \Delta S(\mathbf{x}^{1}) &\mathcal{G} (\mathbf{x}^{N}, \mathbf{x}^{2}) \Delta S(\mathbf{x}^{2}) &\cdots &\mathcal{G} (\mathbf{x}^{N}, \mathbf{x}^{N}) \Delta S(\mathbf{x}^{N})\\
\end{array}
\right)$
},
\end{equation}
and the $\mathcal{G}(\mathbf{x}^i,\mathbf{x}^j)$'s are $3\times3$ matrices of Eq.~(\ref{greens}).
Here $\Delta S(\mathbf{x}^{i})$ is the area occupied by the mesh element at position $\mathbf{x}^i$, and the indices $i$ = 1, 2, $\cdots$, $N$. The 
matrix $\mathcal{G}$ is typically not sparse, and we must solve $\textrm{rank}(\mathcal{L})=3N$ linear equations to find the force density.
The number of equations we are required to solve increases linearly with the number of nodes on the meshed surface. However, if the surface possesses symmetries, we may reduce the size of the linear system dramatically.

Let us consider a system with helical symmetry,
such as a long helical filament that rotates and translates along its axial direction in a Stokes flow.  
A segment of the filament is shown in Fig.~\ref{fig:1}. The 
axis of the helix is along $x_3$. Consider an arbitrary circumference $C_0$ of 
the filament, defined by intersecting
the surface by the cross-section normal to the body centerline. Note that these cross-sections at different points of the helix are related by rotations about $x_3$, since, as we review below, the normal and binormal vectors of the Serret-Frenet frame~\cite{Powers2010} lie in these cross-sections, and the Serret-Frenet frame of a helix rotates about  $x_3$ as the arclength increases.
Now suppose $\mathbf{f}(\mathbf{x})$ is the force-density vector  located at an arbitrary position $\mathbf{x}$. Because of the helical symmetry, there exists a point $\mathbf{x}_C$ on the circumference $C_0$ such that the force density $\mathbf{f}(\mathbf{x}_C)$ satisfies
\begin{equation}
\mathbf{f}(\mathbf{x})=\mathcal{R}^{3} (\Delta \varphi)\cdot \mathbf{f}(\mathbf{x}_C),\label{eq:rot1}
\end{equation}
where $\mathcal{R}^{i}$ 
is rotation 
about axis $x_i$ by $\Delta \varphi$. 
The angle 
$\Delta \varphi$ is given by 
\begin{equation}
\Delta \varphi = \varphi (\mathbf{x}_C)-\varphi (\mathbf{x}),
\end{equation}
where $\varphi(\mathbf{x})$ is the 
angle between  the $x_1$ axis and the projection of the normal $\mathbf{n}(\mathbf{x})$ onto the $x_1$-$x_2$ plane (Fig. 1).
The dashed curve in Fig.~\ref{fig:1} shows a contour $C_{\mathrm{x}_C}$ that goes through $\mathbf{x}$ and exhibits the helical symmetry. Force densities along this contour can always be computed from the force-density vector at $\mathbf{x}_C$, using equation Eq.~(\ref{eq:rot1}). Moreover, the angle $\varphi$ is the phase of the helical wave along the body-centerline, and varies by $2\pi$ over one period of the filament. This more generalized notion of $\varphi$ is useful for determining $\Delta \varphi$ when the surface normal is parallel to the axial direction. By applying this strategy, the force densities over the entire helical surface can be expressed as those on the circumference $C_0$, and the boundary-integral formulation can be written as 
\begin{equation}
u_j(\mathbf{x})=-\frac{1}{8\pi\mu}\int_D \textrm{d}S_{\mathbf{x'}} \mathcal{R}^3_{ik}\left(\varphi(\mathbf{x}')-\varphi(\mathbf{x}_C')\right) f_k(\mathbf{x}'_C)\mathcal{G}_{ij}(\mathbf{x}', \mathbf{x}). \label{bi-	linear}
\end{equation}
Since the flow field, $\mathbf{u}(\mathbf{x})$, also satisfies the helical symmetry, the only flow velocities we need to consider are those distributed on the circumference. Thus the integral equation for $f_k(\mathbf{x}_C)$ is reduced to one dimension:
\begin{equation}
u_j(\mathbf{x}_C)=-\frac{1}{8\pi\mu}\int_{C_0} \textrm{d}l\ f_k(\mathbf{x}'_C)\mathcal{H}_{kj}(\mathbf{x}'_C, \mathbf{x}_C). \label{bi-red}
\end{equation}
Here, the tensor $\mathcal{H}$ is our modified Stokeslet: 
\begin{equation}
\mathcal{H}_{kj}(\mathbf{x}'_C, \mathbf{x}_C)=\int_{C_{\mathbf{x}'_C}} \textrm{d}l_{\mathbf{x'}} \mathcal{R}^3_{ik}\left(\varphi(\mathbf{x}')-\varphi(\mathbf{x}_C')\right) \mathcal{G}_{ij}(\mathbf{x}', \mathbf{x}_C), \label{bi-modstokes}
\end{equation}
where the integration is performed on 
the helical contour $C_{\mathbf{x}'_C}$ that contains the point $\mathbf{x}'_C$. 
Recall that along this contour (such as the dashed curve in Fig.~\ref{fig:1}), 
the force densities are of the same magnitude but have varying orientation. 
The modified Stokeslet depends only on the geometry of the surface, and it can be computed  once the mesh is given.

Equations~(\ref{bi-red}) and~(\ref{bi-modstokes}) form a complete set of equations to solve. The solution to the boundary integral problem is decomposed into two steps: (1) compute $\mathcal{H}$ by performing the integral {in Eq.~(\ref{bi-modstokes}) for each pair of nodes along $C_0$, and (2) solve the linear equations Eq.~(\ref{bi-red}) for $f_k(\mathbf{x}_C)$. 
The number of linear equations is reduced to the number $N_\alpha$ of mesh nodes along a single circumference, for each of the three spatial dimensions.
To solve Eq.~(\ref{bi-red}), we must invert the matrix $\mathcal{H}$, which is  $3 N_\alpha\times3 N_\alpha$. This matrix is much smaller than the $3N\times 3N=3 N_\alpha N_\varphi \times 3 N_ \alpha N_\varphi$ matrix 
$\mathcal{G}$, which must be inverted in the full boundary-element-method case. The spatial resolution along the body centerline 
determines the accuracy of the modified Stokeslet $\mathcal{H}$. 
The computational cost can thus be reduced by orders of magnitudes for the same number of mesh elements. It should be noted that the integrals introduced by Eq.~(\ref{bi-modstokes}) add no additional computational cost when compared to the cost of computing the matrix $\mathcal{G}$, since each element of $\mathcal{G}$ need only be computed once. 

\section{Body-centerline coordinates}

In Sec.~II, we described the general idea 
of how the symmetry of a helical domain can reduce the dimensionality of the boundary-integral method from three to one.
In this section, we implement the idea by introducing a coordinate frame $\{\hat{\mathbf{q}}_1,\hat{\mathbf{q}}_2,\hat{\mathbf{q}}_3\}$ that rotates about ${x}_3$ at the same rate the cross-sections of the helix rotate. In this coordinate system, the components of the force density are the same for every point on the body centerline. 

Figure \ref{fig:2} shows the rotating  frame $\{\hat{\mathbf{q}}_1,\hat{\mathbf{q}}_2,\hat{\mathbf{q}}_3\}$. This frame is the Serret-Frenet frame. We briefly summarize the properties of this frame for a helical curve. 
The filament body-centerline $\mathbf{x}^0$ follows the path 
$\mathbf{r}(s)=(R\cos\varphi(s),R\sin\varphi(s),x_3^0(s))$, where
$s$ is the arc length along the body centerline. If the pitch of the centerline is $\lambda$, then $\Gamma^2=\lambda^2+4\pi^2R^2$ is the arc length of one pitch of the centerline. Defining the pitch angle $\theta$ to be the angle between the centerline tangent $\hat{\mathbf{q}}_3$ and the $x_3$ axis, we find $\sin\theta=2\pi R/\Gamma$ and $\cos\theta=\lambda/\Gamma$, and therefore ${\mathrm d}\varphi/{\mathrm d}s=2\pi/\Gamma=(\sin\theta)/R$. Thus the centerline tangent vector is given by $\hat{\mathbf{T}}=\mathrm{d}\mathbf{r}/\mathrm{d}s=(-\sin\theta\sin\varphi,\sin\theta\cos\varphi,\cos\theta)$.  The normal to the curve is given by the direction of $\mathrm{d}\hat{\mathbf{T}}/\mathrm{d}s$, or $\hat{\mathbf{N}}=(-\cos\varphi,-\sin\varphi,0)$, and the binormal is $\hat{\mathbf{B}}=\hat{\mathbf{T}}\times\hat{\mathbf{N}}$.  The moving frame is hence defined by $\{\hat{\mathbf{q}}_1,\hat{\mathbf{q}}_2,\hat{\mathbf{q}}_3\}\equiv\{\hat{\mathbf{N}},\hat{\mathbf{B}},\hat{\mathbf{T}}\}$. 

Note that the Serret-Frenet frame $\{\hat{\mathbf{q}}_1,\hat{\mathbf{q}}_2,\hat{\mathbf{q}}_3\}$ is obtained by rotating the space-fixed frame $\{\hat{\mathbf{x}}_1,\hat{\mathbf{x}}_2,\hat{\mathbf{x}}_3\}$ by $\varphi$ about the $x_3$ axis, and then rotating the resulting frame by $\theta$ about 
$\hat{\mathbf{q}}_1=\hat{\mathbf{x}}^\prime_1=\hat{\mathbf{x}}_1\cos\varphi+\hat{\mathbf{x}}_2\sin\varphi$. 
The angles $\varphi$ and $\theta$ are Euler angles, and we may relate coordinates in the two frames by
\begin{figure}
\begin{center}
\includegraphics[width=0.8\textwidth]{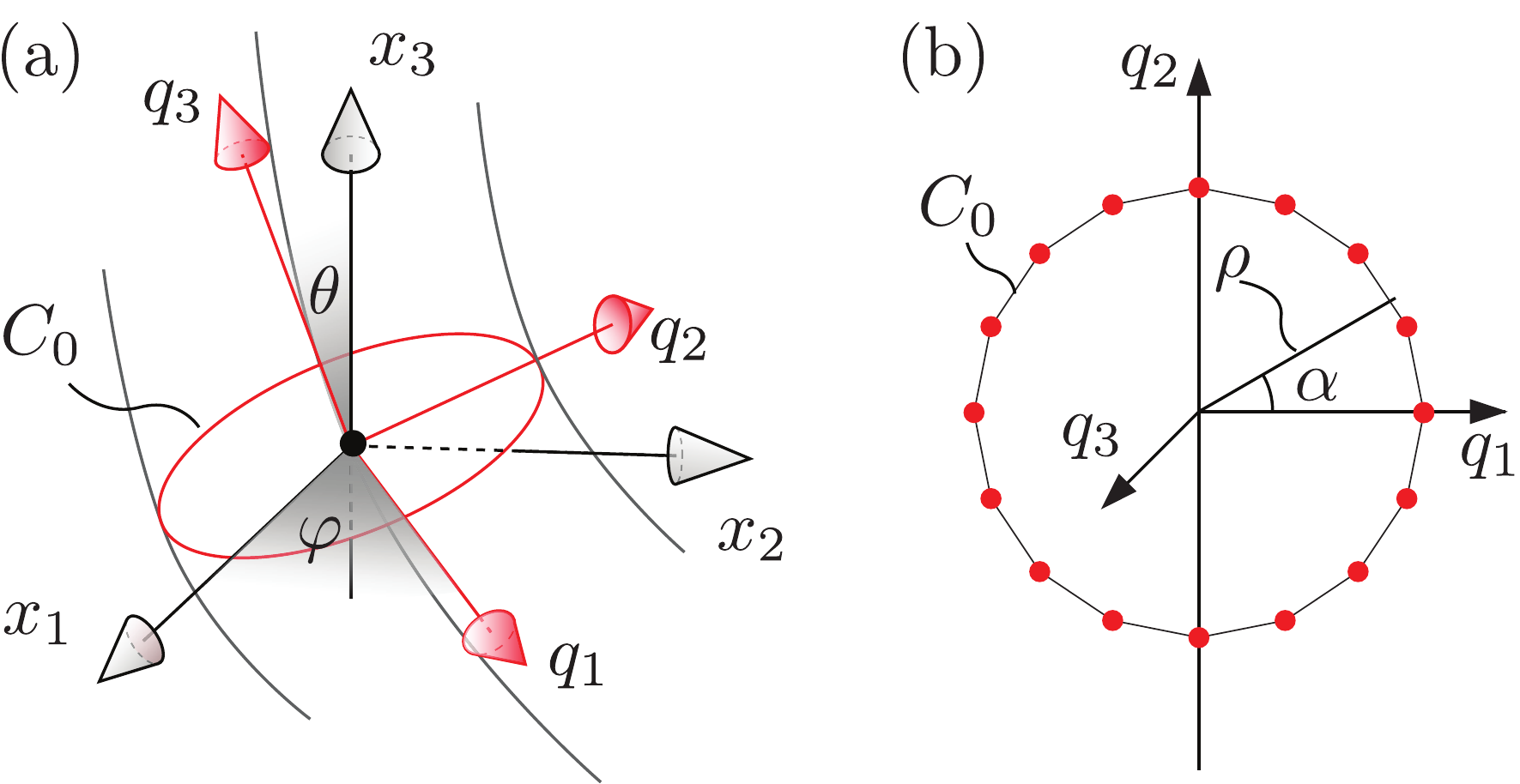}
\caption{(Color online.) Body-centerline coordinate system. (a) A coordinate system that follows the body-centerline of a filament $\left\{q_i\right\}$ is given by two Euler angles $\theta$, $\varphi$. Here $q_3$ is tangential to the body-centerline, and the circumference $C_0$ is 
a circle and normal to $q_3$. (b) A view of the circumference $C_0$ in the body-centerline frame. Radius $\rho$ and angle $\alpha$ are the associated polar coordinates in the body-centerline frame. The dots show the grid points on a single circumference. \label{fig:2}}
\end{center}
\end{figure}
\begin{equation} 
\label{eqtrsflab}
q_i-q^0_i=\mathcal{R}^1_{ij}(\theta)\mathcal{R}^3_{jk}(\blbr{-}\varphi) (x_k-x^0_k), 
\end{equation}
where $\textbf{q}^0=(q^0_1, q^0_2, q^0_3)=(0,0,s)$ 
describes the body-centerline in the new coordinate system. The 
operation in~(\ref{eqtrsflab}) is a change of basis from the space-fixed frame to the body-fixed frame that accounts properly for the different choices of origins, $\textbf{q}^0$ and $\textbf{x}^0$. The rotation operators $\mathcal{R}^1$ and $\mathcal{R}^3$ are
\begin{equation}
\mathcal{R}^1(\theta)=\left(
\begin{array}{ccc}
1 & 0 & 0 \\
0 & \cos\theta & \blbr{-}\sin\theta\\
0 & \sin\theta & \cos\theta\\
\end{array}
\right),
\quad
\mathcal{R}^3(\varphi)=\left(
\begin{array}{ccc}
\cos\varphi & \blbr{-}\sin\varphi & 0\\
\sin\varphi & \cos\varphi & 0\\
0 & 0 & 1\\
\end{array}
\right).
\end{equation}

Figure~\ref{fig:2} (b) shows a cross-section of helical filament in this new coordinate $\mathbf{q}$. It should be noted that such coordinate system is not unique. For a round filament, this cross-section is circular if the selected coordinate systems orient along the body-centerline.

Now let $U$, $f_i$, and $T_{ij}$ denote a scalar, vector, and tensor, respectively, in the Cartisian coordinates $x_i$, and let  $\tilde U$, $\tilde{f}_i$, and $\tilde{T}_{ij}$ denote the same in the moving coordinate system $q_i$.  The relations between these quantities in the two coordinate systems are given by
\begin{eqnarray}
\tilde{U}(\mathbf{q (x)})&=&{U}(\mathbf{x})\\
\tilde{f_i}(\mathbf{q (x)})&=&\mathcal{R}^1_{ij}(\theta) \mathcal{R}^3_{jk} (\blbr{-}\varphi) {f_k}(\mathbf{x}) \label{bi-trsf}\\
\tilde{T}_{ij}(\mathbf{q(x)}) &=& \mathcal{R}^1_{im}(\theta)\mathcal{R}^3_{mk}(\blbr{-}\varphi) T_{kl} (\mathbf{x})\mathcal{R}^{3}_{ln}(\blbr{-}\varphi)\mathcal{R}^1_{nj}(\theta), 
\end{eqnarray}
\blbr{where angle $\theta$ is constant and $\varphi$ is a linear function of $q^0_3=s$.} 
In the rotating coordinate frame, the helical symmetry eliminates the dependence on $q_3$:
\begin{equation}
\frac{\partial \tilde{U}(\mathbf{q})}{\partial q_3}=0, \quad \frac{\partial \tilde{f}_i(\mathbf{q})}{\partial q_3}=0, \quad \frac{\partial \tilde{T}_{ij}(\mathbf{q})}{\partial q_3}=0. \label{mani2d}
\end{equation}
By substituting the force densities and velocity fields in Eq.~(\ref{bi-cartesian}) by Eq.~(\ref{bi-trsf}), 
By using Eq.~(\ref{bi-cartesian}) to express the force density and velocity in terms of the coordinates $q_i$, 
the boundary integral formula Eq.~(\ref{bi-trsf}) becomes
\begin{equation}
\tilde{u}_i(\mathbf{q})=\frac{1}{8\pi\mu}\int_{C_0} \textrm{d}l_{\mathbf{q}'} \det \mathcal{J}\tilde{f}_j(\mathbf{q}')\tilde{\mathcal{H}}_{ji}(\mathbf{q}',\mathbf{q}), \label{bi-bc01}
\end{equation}
where $\mathcal{J}$ is the 2-D Jacobian arising from the change of coordinates from $x_i$ to $q_i$. To close the problem, 
$\mathbf{q}$ and $\mathbf{q}'$ in Eq.~(\ref{bi-bc01}) are both constrained to lie on the circumference $C_0$, i.e., $q_3=q_3'=s_C$. The modified Stokeslet $\tilde{\mathcal{H}}$ is 
\begin{equation}
\tilde{\mathcal{H}}_{ji}(\mathbf{q}', \mathbf{q})=\int_{-Q/2}^{Q/2} \textrm{d}q_3' \mathcal{P}_{jk}(\theta(q_3'), \varphi(q_3'))\mathcal{G}_{km}(\mathbf{x}(\mathbf{q'}), \mathbf{x(q)})\mathcal{P}^{-1}_{mi}(\theta(q_3), \varphi(q_3)), \label{bi-bc02}
\end{equation}
where $Q$ is the total arc length of the filament,  and the tensor $\mathbf{\mathcal{P}}$ 
is the rotation operator that takes the space-fixed frame to the moving frame,
\begin{equation}
\mathcal{P}_{ij} (\theta, \varphi)=\mathcal{R}^1_{ik}(\theta ) \mathcal{R}^3_{kj}(\blbr{-}\varphi).
\end{equation}

Equations (\ref{bi-bc01}) and (\ref{bi-bc02}) may be further simplified using polar coordinates $\rho$ and $\alpha$, where  $q_1=\rho \cos \alpha$, and $q_2=\rho \sin\alpha$. Since $\rho=a$ along the circumference $C_0$, Eq.~(\ref{bi-bc01}) simplifies to
\begin{equation}
\tilde{u}_i(\mathbf{q}(\alpha))=\frac{1}{8\pi\mu}\int_{C_0} \textrm{d}\alpha'
a \tilde{f}_j^J(\mathbf{q}(\alpha'))\tilde{\mathcal{H}}_{ji}(\mathbf{q}(\alpha'),\mathbf{q}(\alpha)),
\label{bi-bc01_2} 
\end{equation}
where the vector $\tilde{\mathbf{f}}^J (\mathbf{q})$ is defined as $\tilde{\mathbf{f}}^J (\mathbf{q})\equiv \det \mathcal{J} \tilde{\mathbf{f}}(\mathbf{q})$. 
Likewise, 
since $q_3=s$, we can choose the origin of the coordinate $\mathbf{q}$ so that 
\begin{equation}
q_3=\Gamma\varphi/(2\pi),
\end{equation}
where $\Gamma$ is the arc length of the filament within each period. 
Thus Eq.~(\ref{bi-bc02}) can be written as
\begin{equation}
\tilde{\mathcal{H}}_{ji}(\mathbf{q}(\alpha'), \mathbf{q}(\alpha))=\frac{\Gamma}{2\pi}\int_{-\kappa\pi}^{\kappa\pi} \textrm{d}\varphi' \mathcal{P}_{jk}(\theta, \varphi')\mathcal{G}_{km}(\mathbf{x}(\mathbf{\varphi', \alpha'}), \mathbf{x(\varphi_C,\alpha)})\mathcal{P}^{-1}_{mi}(\theta, \varphi_C), \label{bi-bc02_2}
\end{equation}
where $\kappa$ is the number of helical pitches (or the wave number), and $\varphi_C$ is phase of the body-centerline at $C_0$. To assure that the helical symmetry is a valid approximation, we use the convention that $C_0$ is located in the middle of the filament and $\varphi_C=0$.

\section{Numerical simulations of a helical swimmer}

In this section we implement the boundary-integral method for a rotating helix with many turns. Now that we have the boundary-integral 
equations~(\ref{bi-bc01_2}) and (\ref{bi-bc02_2}), the next step is to discretize them. We use an rectangular grid with $N_\varphi\times N_\alpha$ points per helical pitch, so that the total number of grid points is $\kappa\times N_\varphi\times N_\alpha$. Using the trapezoidal rule for the integrals, we find
\begin{eqnarray}
\tilde{u}_i(\alpha_m)&=&\frac{1}{8\pi\mu}\sum_{l=1}^{N_\alpha}\Delta\alpha \rho (\alpha_l)  \sum_{j=1}^{3}\tilde{f}_j^J(\alpha_l)\tilde{\mathcal{H}}_{ji}(l,m),\label{bi-bc01_3}\\
\tilde{\mathcal{H}}_{ji}(n, m)&=&\frac{\Gamma}{2\pi}{\sum_l}' w(l,n) \Delta \varphi  \sum_{k=1}^{3}\sum_{k'=1}^{3} \mathcal{P}_{jk}(\theta, \varphi_l)\mathcal{G}_{kk'}(\mathbf{x}(\varphi_l, \alpha_n), \mathbf{x}(0,\alpha_{m}))\mathcal{P}^{-1}_{k'i}(\theta, 0) \nonumber\\
&&+ \delta_{nm} \mathcal{E}_m(\Delta \varphi, \Delta \alpha),\label{bi-bc02_3}
\end{eqnarray}
where $w(l, n) \in \{ 1/4$,  $1/2$, $3/4$, $1$\} is the weight function for the trapezoidal rule. Note that the sum omits the terms where the integrand is singular. These terms, with $\varphi_l=\varphi_0$ and $m=n$, are accounted for in $\mathcal{E}_m(\Delta \varphi, \Delta \alpha)$. To ensure good accuracy, the contributions to the integral from the singular parts of the integrand are computed analytically; see Appendix~B. For a given rigid body with prescribed velocity, our task is to solve Eqs.~(\ref{bi-bc01_3}--\ref{bi-bc02_3}) for the force per unit length $\tilde{f}^J_j(\alpha_l)$ by inverting the $3N_\alpha\times3N_\alpha$ matrix $\tilde{\mathcal H}_{ji}(l,m)$. Note that without exploiting the helical symmetry, we would have to invert a $3N_\alpha N_\varphi\times 3N_\alpha N_\varphi$ matrix.

We consider two cases: (1) a tethered helix, which rotates but is prevented from translating, and (2) a swimming helix, which is subject to zero net force. Note that in both cases an external torque drives the motion. For the tethered helix, the velocity is given by 
$\mathbf{u}=\Omega \hat{\mathbf{x}}_3\times \mathbf{x}$. In the helical coordinate system $\mathbf{q}$, the helix velocity takes the form $\tilde{u}_i(\alpha_m)=\mathcal{R}^1_{ij}(\theta)  u_j(\alpha_m)$.
\begin{figure}
\begin{center}
\includegraphics[width=1\textwidth]{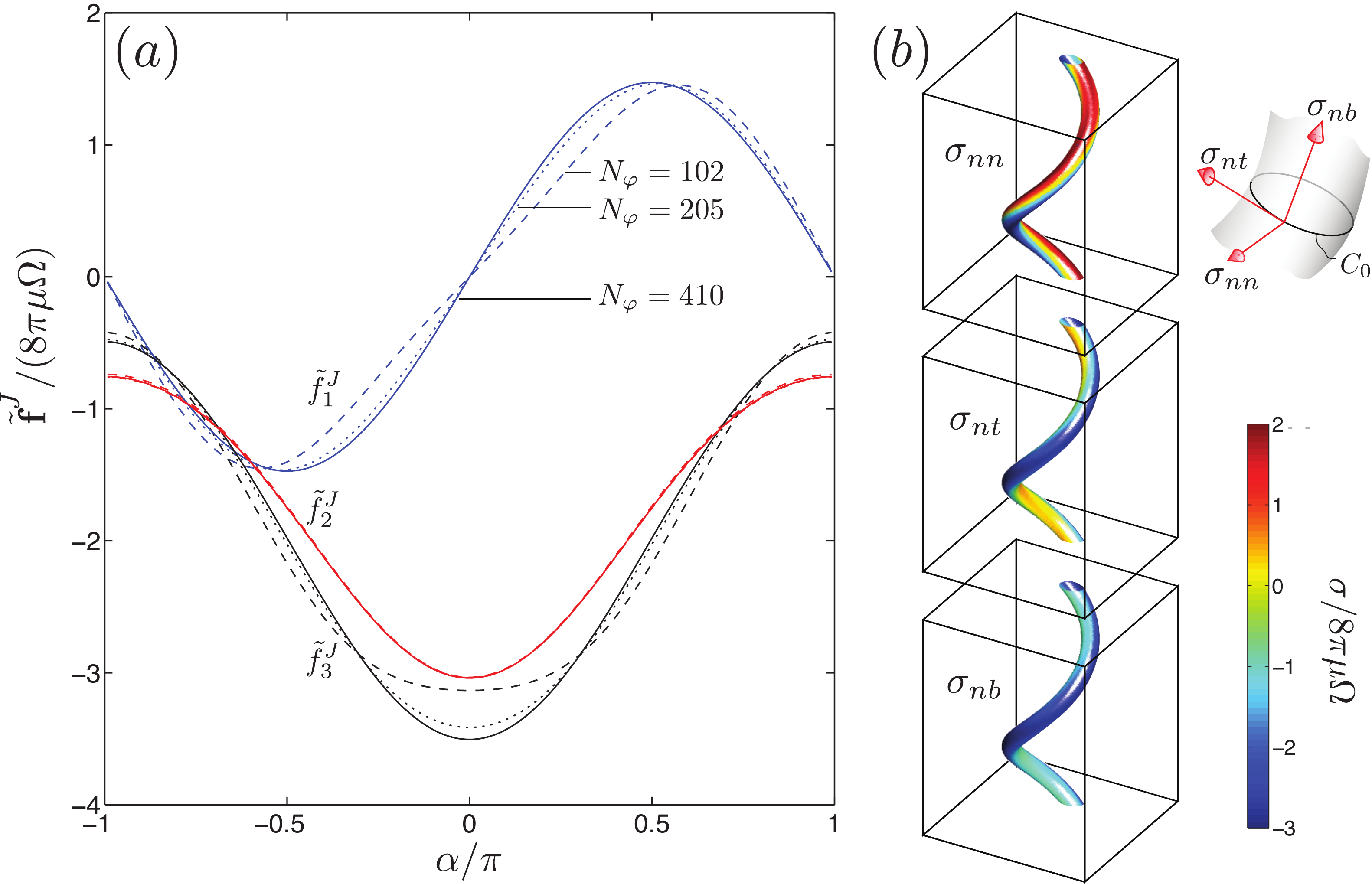}
\caption{(Color online.) Force density on a rotating tethered helix 
with $a/\Gamma=0.026$, $\theta=\pi/4$,  $\kappa=40$, and $N_\alpha=128$. (a) The force density 
$\tilde{\mathbf{f}}^J(\alpha)$ along a circumference approaches the exact value with a 
second-order convergence with respect to $N_\varphi$. 
(b) A three dimensional distribution of the stress tensor $\sigma$ is computed
from $\tilde{\mathbf{f}}^J(\alpha)$ using the helical symmetry. The inset illustrates the direction of three components of $\sigma$: $\sigma_{nn}$, $\sigma_{nt}$, $\sigma_{nb}$ are normal, tangential, binormal to the circumference $C_0$, respectively.
\label{fig:3}}
\end{center}
\end{figure}
Figure~\ref{fig:3} shows a typical simulation result 
for the force density 
on a tethered helix with parameters 
$a/\Gamma=0.026$ and $\theta=\pi/4$. To ensure helical symmetry, the number of the pitches is set at $\kappa=40$. 
Given the spatial grids, matrices $\tilde{H}_{ji}(n, m)$ are computed using Eq.~(\ref{bi-bc02_3}). Discretized force densities $\tilde{\mathbf{f}}^J(\alpha_i)$, $i=1, 2 \cdots N_\alpha$, are thus obtained by inverting the linear equation (Eq.~(\ref{bi-bc01_3})). As shown in Fig.~\ref{fig:3}, 
$\tilde{\mathbf{f}}^J(\alpha)$ 
converges as $N_\varphi$ increases. Among 
the three components of the force density,  the component $\tilde{f}_2^J$ converges fastest. The details of the convergence analysis are described in Appendix C. Using the helical symmetry, we reconstruct the force density on the entire helical surface 
(Fig.~\ref{fig:3}(b)). 
The figure shows three components of the stress tensor distribution, ($\sigma_{nn}$, $\sigma_{nt}$, $\sigma_{nb}$), with components along the normal, tangential, and bi-normal directions, respectively (inset of Fig.~\ref{fig:3}(b)). Such high spatial resolution of the stress distribution provides an accurate method in computing many mechanical features on helical propulsion, such as its net power consumption.

Using the same algorithm, we also calculate the swimming speed of a rotating helix.
For a free swimmer with swimming speed $V_0$, the velocity of points on the helix is
\begin{equation}
\mathbf{u}=\Omega \hat{\mathbf{x}}_3\times \mathbf{x}+V_0 \hat{\mathbf{x}}_3.
\end{equation}
The swimming speed $V_0$ is determined by the condition of vanishing axial force,
\begin{equation}
\sum_{l=1}^{N_\alpha} \left(\tilde{f_2}^J(\alpha_l) \sin\theta + \tilde{f_3}^J(\alpha_l) \cos\theta \right) = 0.\label{bi-fr}
\end{equation}
The free-swimming speed $V_0$ exhibits similar convergence properties as those of the force densities (see Appendix C). As also demonstrated in Appendix C, by properly removing the lowest order of the numerical errors, a robust third-order convergence can be 
obtained. 
 
\section{Comparison with experiments and theories}

In this section we compare the results of our boundary-integral method for a swimming helix with experiments and the predictions of resistive-force theory and two different slender-body theories.

\subsection{Experimental system}

The experimental system is described in our previous work \cite{Liu:PNAS2011}: a  helical filament 
a few centimeters long ($L\sim 10$ cm) and millimeters wide ($R \sim1$ mm) simultaneously 
rotates and translates along its axial direction in a viscous fluid. 
To achieve force-free swimming, we fix its rotation rate $\Omega$, and vary the translation speed $V$ 
until the net  hydrodynamic force $F_\textrm{hydro}$ on the helix vanishes. The translation speed $V_0$ corresponding to 
$F_\textrm{hydro}=0$ is the free-swimming speed. The fluid 
is a high molecular weight silicone oil with 
kinematic viscosity $\mu \approx 10^3$\,St. In the regime of our experiment, the fluid is 
Newtonian, with the viscosity almost independent of the shear rate. For 
the typical rotation rate, $\Omega \sim 10$\,rad/s, the Reynolds number $\textrm{Re}=\Omega R^2/\mu \sim 10^{-4}$, and the inertia of the fluid is thus negligible. As the helix is inserted in the fluid, we find that the force-free swimming speed, $V_0$, does not vary with time once about one helical turn has been immersed in the fluid. Figure~\ref{fig:5} shows the results of our measurements for two different filament thicknesses and a few different pitch angles~\cite{Liu:PNAS2011}.

\subsection{Resistive force theory}

\begin{figure}
\begin{center}
\includegraphics[width=.5\textwidth]{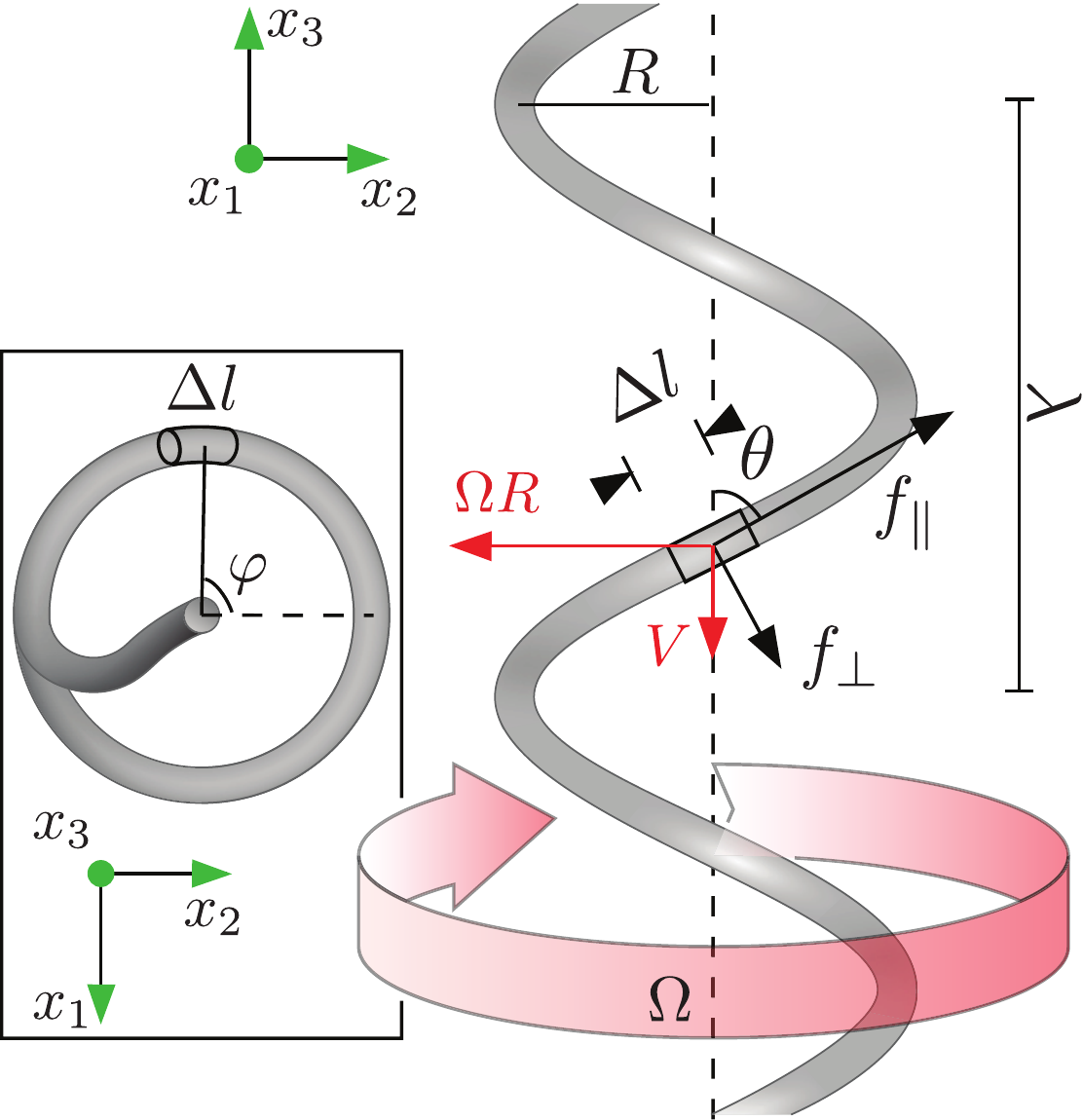}
\caption{(Color online.) Helix geometry. The helix rotates with rate $\Omega$ and advances at constant speed $V$ along its axis of symmetry. 
\label{fig:4}}
\end{center}
\end{figure}

Resistive-force theory, or local drag theory, is the simplest approximation for computing the force on a thin body at low Reynolds number~\cite{Gray:1955}. The  force per unit length, $\mathbf{f}$, that a small segment of the body  exerts on the fluid is taken to be proportional to the body's local velocity, $\textbf{u}$, relative to the fluid far away, with different proportionality constants  $C_\perp$ and $C_{||}$ for motion perpendicular and parallel to the body centerline (see Fig.~\ref{fig:4}):
\begin{equation}
f_\perp=\mu C_\perp u_\perp, \quad f_{||}=\mu C_{||} u_{||},
\end{equation}
where $\mu$ is fluid viscosity. The linearity of Stokes equations implies that the total force per wavelength that the helix exerts along the $x_3$ axis is $F_3=A V + B R \Omega$, where $A$ and $B$ are constants proportional to viscosity $\mu$. To compute $A$, note that the force per wavelength required to pull a helix along its axis but with $\Omega=0$ is
\begin{equation}
F^\mathrm{drag}_3=\mu V\left(C_{||}\cos^2\theta+C_\perp\sin^2\theta\right)\Gamma.
\end{equation}
Likewise, $B$ is determined by the force required to prevent a rotating helix from translating,
\begin{equation}
F^\mathrm{thrust}_3=-\mu \Omega R\left(C_\perp-C_{||}\right)\cos\theta\sin\theta\Gamma.
\end{equation}
The swimming speed is determined by demanding that the total force vanish, $F_3=F^\mathrm{drag}_3+F^\mathrm{thrust}_3=0$. \blbr{As proposed by Gray and Hancock~\cite{Gray:1955}, the drag coefficients are
$C_{||}=2\pi /(\ln\frac{2\lambda}{a}-\frac{1}{2})$ and $C_\perp=4\pi /(\ln\frac{2\lambda}{a}+\frac{1}{2})$}. 
Adopting for simplicity 
the coefficients in the limit $\lambda\gg a$, 
we find $V_0/(\Omega R) = \sin\theta\cos\theta/(\sin^2\theta+1)$. 
\blbr{Lighthill also proposed a set of drag coefficients~\cite{Lighthill:1976}, 
$C_{||}=2\pi /(\ln\frac{0.18\Gamma}{a})$ and $C_\perp=4\pi /(\ln\frac{0.18\Gamma}{a}+\frac{1}{2})$, 
which were optimized for helical filaments.}
The predictions of resistive-force theory for $V_0/(\Omega R)$  as a function of $\theta$ are shown in Fig.~\ref{fig:5}. Note that the \blbr{curves are} not symmetric about $\theta=\pi/4$\blbr{, even in the limit that $\lambda\gg a$}. The maximum speed is at a pitch angle less than $\pi/4$ because although the thrust is maximized at $\theta=\pi/4$, the drag is minimized at $\theta=0$, since $C_{||}<C_\perp$.

\subsection{Slender-body theory}

We now turn to slender-body theory in which the filament is modeled by a one-dimensional distribution of singular solutions to Stokes equations. The flow at any point on the surface of the filament is given by integrating the contributions to the flow from all the other parts of the filament. Since the filament is represented by a one-dimensional distribution of singular solutions, the accuracy of slender-body theory is controlled by the aspect ratio, $\varepsilon = a/L$, with the error vanishing as $\varepsilon\rightarrow0$. 

\subsubsection{Lighthill's slender-body theory}

Lighthill gave a simple physical derivation for an integral relating the velocity of a point on a filament to the distribution of forces per unit length acting on the filament~\cite{Lighthill:1976,lighthill1996b}.  The singular solutions are point forces, or Stokeslets, and source doublets. Let $\mathbf{r}(s)$ denote the centerline of the filament, where $s$ is arclength. For our helix, 
$\mathbf{r}(s)=\hat{\mathbf{x}}_1R\cos\varphi  +  \hat{\mathbf{x}}_2R\sin\varphi +\hat{\mathbf{x}}_3 \lambda\varphi/(2\pi)$, where $\varphi=2\pi s/\Gamma$ is the polar angle.

Since Stokes flow depends only on the instantaneous velocity of the filament, and since we consider rigid body motion, it is sufficient to consider the position of the helix at only one instant of time. 
Denoting the vector from one point on the helix to another by $\mathbf{X}(s,0)=\mathbf{r}(s)-\mathbf{r}(0)$, Lighthill's slender-body theory formula for the velocity $\mathbf{u}(0)$ of a point on the centerline of the filament is 

\begin{equation}\mathbf{u}(0) = \frac{\mathbf{f}_\mathrm{n}(0)}{4\pi\mu}
+\frac{1}{8\pi\mu}\int_{|\mathbf{X}(s',0)|>\delta}\frac{\mathbb{I}+\hat{\mathbf{X}}(s',0)\hat{\mathbf{X}}(s',0)}{|\mathbf{X}(s',0)|}\cdot\mathbf{f}(s')\mathrm{d}s',\end{equation}
where $\mathbf{f}_\mathrm{n}$ is the part of $\mathbf{f}$ that is normal to the filament centerline, and $\delta$ the short distance cutoff,
$\delta = a\sqrt{\mathrm{e}}/2$~\cite{Lighthill:1976,lighthill1996b}\blbr{, where `e' is the natural exponent}.
Lighthill argued that the errors in his formula can be as small as $\mathcal{O}(\varepsilon)$, and Childress showed that the errors are no worse than $\mathcal{O}(\varepsilon^{1/2})$~\cite{Childress1981}.

It is convenient to parametrize the helix by the angle $\varphi$. The short-distance cutoff corresponds to a cutoff $\varphi_\mathrm{c}$  in the angle, defined by 
$|\mathbf{X}(s(\varphi_\mathrm{c}),s(\varphi=0))|=\delta$.
In our experiments, the pitch angle $\theta$ is changed for a given helix by stretching the wire, which changes the pitch $\lambda$ and radius $R$, but keeps the contour length fixed. As before we denote the contour length of one helical pitch by $\Gamma$. Since $a\ll \Gamma$, the cutoff angle is approximately
$\varphi_\mathrm{c}\approx a\pi\sqrt{\mathrm{e}}/\Gamma$. 
This expression is accurate to three significant figures over the range of $\theta$ that we measure. 

Just as in our resistive-force theory calculation, the force per unit length at a given point $\mathbf{f}$ has only a $\varphi$ component. Since our experiments show that the zero-force swimming speed is independent of immersed length once the immersed length is greater than one or two wavelengths, we may consider an infinite helix for which the magnitude of $\mathbf{f}$ is uniform. Thus, we find
\begin{equation}\frac{V_0}{\Omega R}=\frac{-\cos\theta\sin\theta +\int^\infty_{\varphi_\mathrm{c}}{\mathrm{d}\varphi\left(\varphi\sin\varphi\cos\theta\csc^2\theta\right)/\xi^{3/2}}}{\cos^2\theta+\int^\infty_{\varphi_\mathrm{c}}\mathrm{d}\varphi\csc\theta\left[\left(\cos\varphi\right)/\xi^{1/2}+\left(\sin^2\varphi\right)/\xi^{3/2}\right],}\end{equation}
where $\xi(\varphi, \theta)=4\sin^2(\varphi/2)+\varphi^2 \cot^2\theta$. This integral is readily evaluated numerically. The predictions of Lighthill's formula for $V_0/(\Omega R)$ are shown in 
Fig.~\ref{fig:5} for two different values of $a/\Gamma$. 

\subsubsection{Johnson's slender-body theory}

Johnson gave a more rigorous derivation of slender-body theory~\cite{JOHNSON:1980p98}, building on ideas of Keller and Rubinow~\cite{keller_rubinow1976}. By assuming that the filament has tapered ends, with a radius $r(s) = \varepsilon\sqrt{4s(L-s)}$, Johnson  derived slender-body theory formulas with $\mathcal{O}(\varepsilon^2\log\varepsilon)$ accuracy. The velocity of a point on the filament is broken into local and nonlocal terms,
$\mathbf{v}(0)=
\mathbf{v}_\mathrm{local}+\mathbf{v}_\mathrm{nonlocal}$, where 
\begin{equation}\mathbf{v}_\mathrm{local}(0)=\frac{1}{8\pi\mu}\left[-\log\left(\varepsilon^2\mathrm{e}\right)\left(\mathbb{I}+\hat{\mathbf{s}}\hat{\mathbf{s}}\right)+2\left(\mathbb{I}-\hat{\mathbf{s}}\hat{\mathbf{s}}\right)\right]\cdot\mathbf{f}(0),\end{equation}
with $\hat{\mathbf{s}}=\mathrm{d}\mathbf{r}/\mathrm{d}s=\hat{\mathbf{T}}$ is the local tangent vector, and
\begin{equation}\mathbf{v}_\mathrm{nonlocal}(0)=\frac{1}{8\pi\mu}\int\left[\frac{\mathbb{I}+\hat{\mathbf{X}}\hat{\mathbf{X}}}{|\mathbf{X}|}\cdot\mathbf{f}(s')-\frac{\mathbb{I}+\hat{\mathbf{s}}\hat{\mathbf{s}}}{|s-s'|}\cdot\mathbf{f}(0)\right]\mathrm{d}s'.\end{equation}
For the helix,
$V_0/(\Omega R)=C_1/C_2$, where
\begin{eqnarray}C_1&=&-\cos\theta\sin\theta\left[2+\log\left(\varepsilon^2\mathrm{e}\right)\right]\nonumber\\&+&2\int_0^{L/2}\left[\frac{\varphi\sin\varphi\cos\theta\csc^2\theta}{\xi^{3/2}}-\frac{\cos\theta\sin\theta}{\varphi}\right]\mathrm{d}\varphi\end{eqnarray}
and
\begin{eqnarray}&C_2&=-\log\left(\varepsilon^2\mathrm{e}\right)\left(1+\sin^2\theta\right)+2\cos^2\theta\nonumber\\&+&2\int_0^{L/2}\left[\left(\frac{\cos\varphi}{\xi^{1/2}}+\frac{\sin^2\varphi}{\xi^{3/2}}\right)\csc\theta - \frac{1+\sin^2\theta}{\varphi}\right]\mathrm{d}\varphi.\end{eqnarray}

The expressions  $C_1$ and $C_2$ are readily evaluated numerically. 
The predictions of Johnson's slender-body theory for force-free swimming helices are also plotted in Fig.~\ref{fig:5}. 

\subsection{Comparison}

\begin{figure}
\begin{center}
\includegraphics[width=0.98\textwidth]{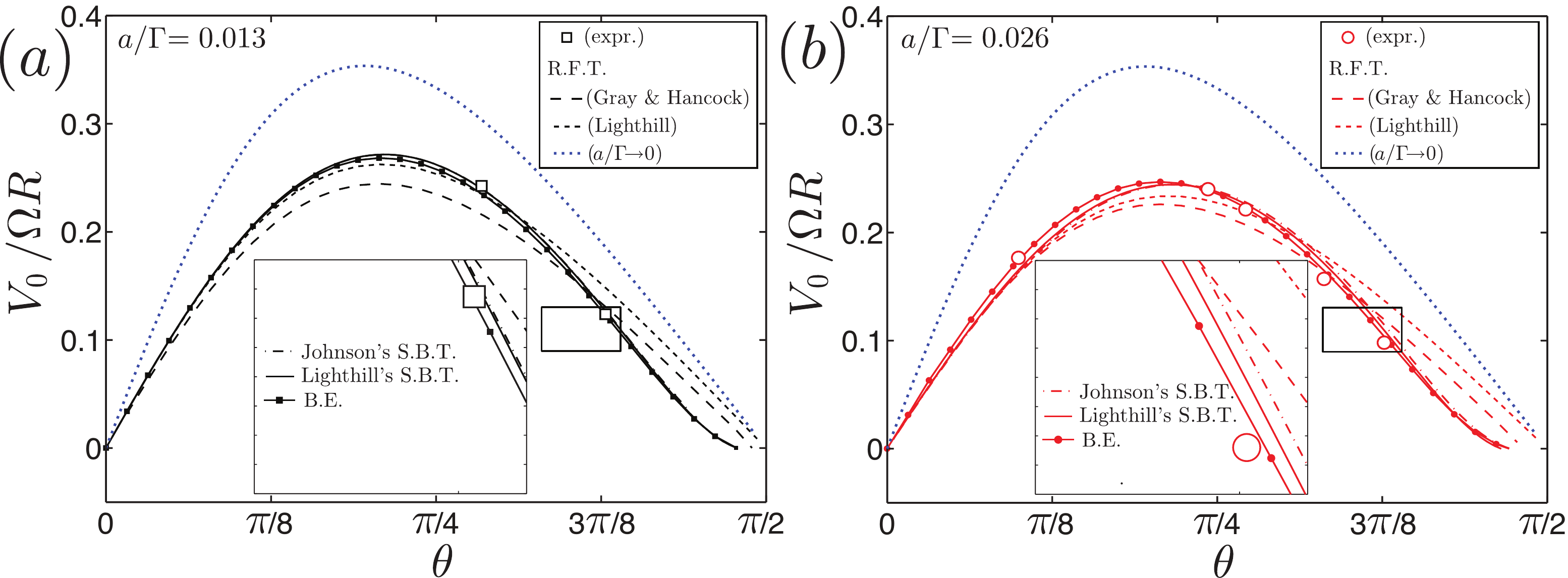}
\caption{(\blbr{Color online.) Comparison of the boundary-integral results with experimental measurements and theoretical predictions for the normalized swimming speed $V_0/\Omega R$ of helices of various pitch angles $\theta$ and two different aspect ratios, (a) $a/\Gamma=0.013$, and (b) $a/\Gamma=0.026$. The graphs show experimental measurements (circles and squares) and the predictions of resistive-force theory (R.F.T.) \blbr{in the limit of $a/\Gamma\rightarrow0$, Gray and Hancock's theory, Lighthill's approximations for finite $a$}, Lighthill and Johnson's slender-body theories, and the boundary-element method with reduced dimension (B.E.). The inset shows a zoom-in view of the rectangular window. 
Adapted from~\cite{Liu:PNAS2011}.} 
\label{fig:5}}
\end{center}
\end{figure}

The simulation results from our modified boundary-element method agrees best with the experimental measurements. Although the resistive-force theory with $a/\Gamma\rightarrow0$ captures the qualitative trend of the dependence of $V_0/(\Omega R)$ on $\theta$, it is not very accurate even for filaments \blbr{with $a/\Gamma=0.013$} [Fig~\ref{fig:5}]. \blbr{The results are much improved when taking into account the finite $a$, by using Gray and Hancock's or Lighthill's drag coefficients. The agreement with Lighthill's version is noticeably better  at small pitch angles for the two groups of filament thicknesses we test. However, these resistive force theories give qualitatively incorrect behavior at larger pitch angles (e.g., $\theta \gtrsim 0.3 \pi$), since the distance between adjacent pitches becomes shorter.} The inaccuracy arises because resistive-force theory does not properly account for the hydrodynamic interactions between different parts of the helix\blbr{~\cite{Rodenborn:2013}}.

For the boundary-element method, the number of helix pitches being simulated is large enough, e.g., $\kappa=20$, so that the free-swimming speed is already independent on the detailed selection of $\kappa$. Note also that there are walls in the experiment, but the simulations and theories do  not account for the wall; apparently the wall is far enough from the helix to have no effect on the speed. Despite the different error estimates, the results from both slender body theories (Lighthill's and Johnson's) are virtually identical. Note that the slope of the $V_0/(\Omega R)$ \textit{vs.} $\theta$ [Fig.~\ref{fig:5}] curve is different for the asymptotic resistive force theory and the slender-body theories (or boundary-element technique) near $\theta=\pi/2$. The difference arises because the asymptotic resistive force theory does not account for the thickness of the filament; sufficiently close to $\theta=\pi/2$, a helix of nonzero thickness and many pitches long will intersect itself. 
For this reason, and because this regime is not physically relevant, we do not carry out our slender-body calculations and boundary-element simulations very close to $\theta=\pi/2$.

\blbr{
\section{Other applications}

Our technique is useful not only for infinite rigid helical filaments with circular cross section. 
It can be applied to many other 
geometries, such as helices with non-circular cross section, helices confined to circular tubes, and non-rigid bodies with helical symmetry. With minor modification, we can even apply our technique to situations that do not exhibit perfect helical symmetry, such as finite-length helices or bodies with non-uniform geometry.  Two examples of these extensions are presented here.

\subsection{Confined geometry and non-rigid body}
First we consider swimming in confined geometry. Noting that a long cylinder is a special case of a helix with pitch angle $\theta=0$, we apply helical symmetry to the study of a helical swimmer in a co-axial cylindrical tube [inset of Fig.~\ref{fig:6}(a)]. The force densities distributed on the helical filament and the cylindrical wall are mapped to those on two circumferences along the above two surfaces, respectively. The dimensional reduction inherent in our technique allows us to study situations when the confining wall is very close to the filament surface without significantly increasing the number of grid points. For example, when the radius, $A$, of the confining tube is comparable to the minimum tube radius, $A_\textrm{min}=R+a$, the number of grid points required is comparable to that required for an unconfined helix and much smaller than the number of grid points required to accurately simulate a helix in a tight-fitting tube using a conventional boundary-element technique. Figure ~\ref{fig:6}(a) shows that, at
a given rotation rate $\Omega$, the free-swimming speed $V_0$ almost increases monotonically with decreasing $A$. 

Besides rigid bodies,  our technique can also be extended to deforming structures by keeping the double-layer potential in Eq.~\ref{eq:bi}. For instance,  a transverse helical wave can propagate along the  body and lead to motility. In rigid-body motion, all points of the helical filament rotate about the axis of the filament. For a helical wave, the filament deforms at every point, and the cross-sections of the filament do not rotate about the filament centerline as the deformation progresses. However, the centerline of a filament carrying a helical traveling wave of frequency $\Omega$ rotates about the helix axis with rotation frequency $\Omega$.
%
%
%
An example of swimming motility by such a helical wave is shown in Fig.~\ref{fig:6}(a), where we see that $V_0/(\Omega R)$ vs. $A/A_{\mathrm {min}}$ is almost identical to the rigid-rotation case. This is reasonable since the filament here is still extremely thin, i.e., $a/\Gamma = 0.013$. 

\begin{figure}
\begin{center}
\includegraphics[width=0.9\textwidth]{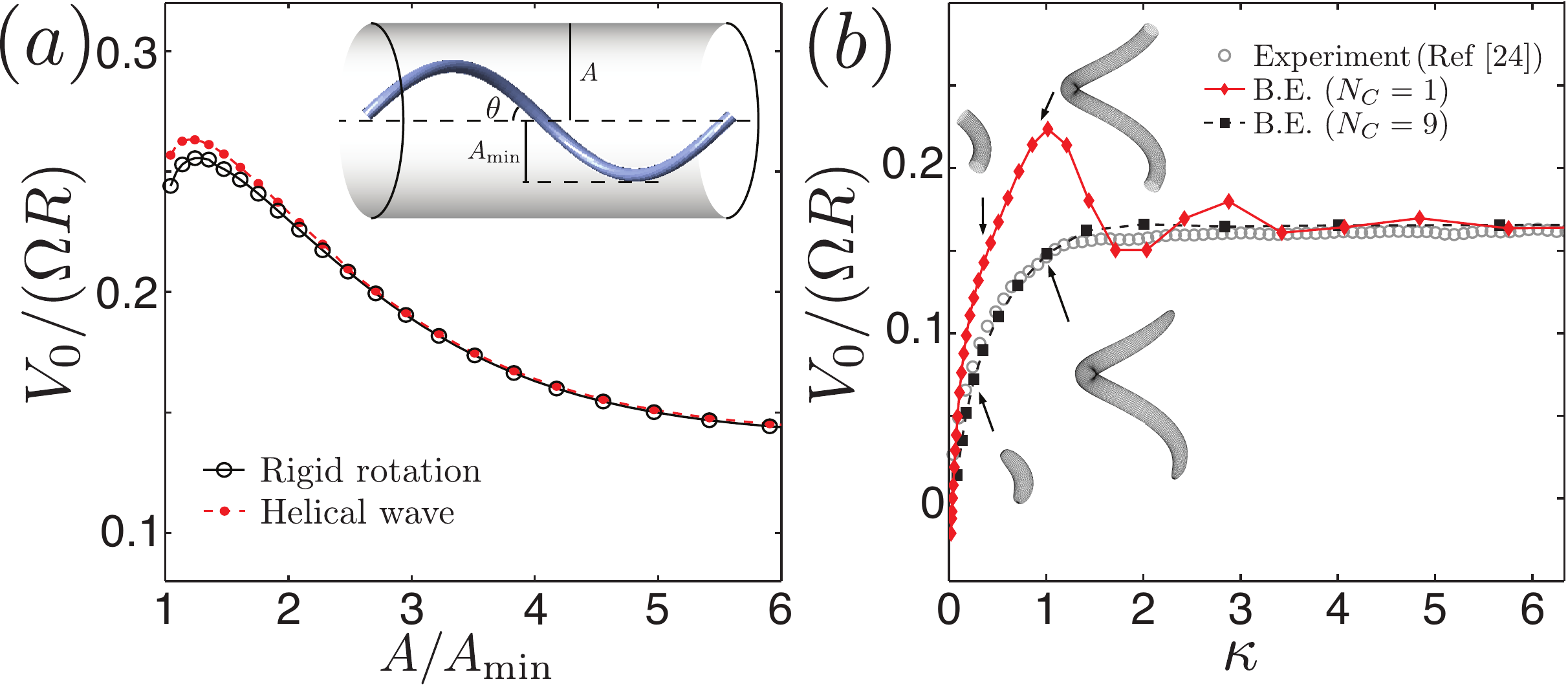}
\caption{\blbr{(Color online.) (a) Normalized free-swimming speed $V_0/\Omega R$ of a helical filament in a tube of radius $A$ due to rigid rotation `$\circ$' and transverse wave `$\bullet$'. Here, the geometry of helical filament is given by $\theta=0.16\pi$, and $a/\Gamma = 0.013$. (b) Effect of finite length on helical swimming using linear interpolation [Eq.~(\ref{eq:interp})] with $N_C=9$ and $N_\varphi=32$, and comparison with the experiment. The result obtained by enforcing full helical-symmetry ($N_C=1$) is also shown as a comparison.}
\label{fig:6}}
\end{center}
\end{figure}

\subsection{Non-uniform geometry}
}
\blbr{As shown previously, we map the force densities on the entire helical surface to those on a single circumference either around the filament or around the confining structure.} This mapping is no longer valid for short, finite-length filaments, where end effects
are important, e.g., when 
$\kappa <1$. 
Nevertheless, the method can still be of value, and in
such situations, instead of mapping the force densities to 
a single circumferences, it is natural to map them to 
a few circumferences spaced along the length of the filament.
For this case, we introduce 
interpolation operators, and the orthogonal mapping in Eq.~(\ref{eq:rot1}) becomes 
\begin{equation}
\mathbf{f}(\mathbf{x})=\sum_{i=0}^{N_C-1}h(\mathbf{x}, \mathbf{x}_{C_i})\cdot \mathcal{R}^{3} (\varphi(\mathbf{x})-\varphi(\mathbf{x}_{C_i}))\cdot \mathbf{f}(\mathbf{x}_{C_i}), \label{eq:interp}
\end{equation}
where $N_C$ is the number of circumferences $\{C_i\}$ onto which the force densities are mapped, and $h(\mathbf{x}, \mathbf{x}_{C_i})$ is the weight function due to interpolation. The size of the linear equations to solve now becomes $3 N_C \times  N_\alpha$. 
\blbr{When $N_C=1$, the formula above (Eq.~(\ref{eq:interp})) reduces to the case with full helical symmetry (Eq.~(\ref{eq:rot1})).} 
Our preliminary studies show that even if $N_C$ is still much less than $\kappa N_\varphi$ (e.g., $N_C=9$ in Fig.~\ref{fig:6}), we obtain 
reliable results due to the approximate helical symmetry. The computational cost is thus much reduced. \blbr{Figure \ref{fig:6}(b) shows an example of such an application. We study the length-dependence of the motility of a helical filament $V_0/\Omega R$, and compare the results with that obtained from the experiments as described in our previous work~\cite{Liu:PNAS2011}. Here, the filament has a finite length $\kappa \Gamma$ and is modeled as an elongated spheroid shape with the radius of its cross-section $r(s) = \varepsilon\sqrt{4s(\kappa \Gamma-s)}$. The aspect ratio $\varepsilon$ is selected such that the mean radius satisfies $\frac{1}{\kappa \Gamma}\int_0^{\kappa\Gamma}r(s) \textrm{d}s  =a$. In both the experiments and the numerical simulation with $N_C=9$, the free-swimming speed $V_0$ saturates at $\kappa \gtrsim 1$. We also validate here the use of helical symmetry for finite-length helices by letting $N_C=1$. The error due to the enforced helical symmetry oscillates but eventually vanishes at $\kappa \gtrsim 6$. Detailed analysis and further applications of this extension to non-uniform body shape will be reported in a separate work.}

\blbr{
\section{Discussion}
We have shown that by exploiting helical symmetry, we are able to reduce a two-dimensional boundary-integral method to a one-dimensional method, without throwing away the detailed small-scale structure, such as the finite thickness of the flagellum and its distance to the surrounding structures.
Our strategy is ideal for situations with symmetry that persists in the entire fluid flow and structure such as the flow generated by an infinitely long rotating helical filament. Nevertheless, in cases such as our experiment, with $\kappa\gtrsim1$, the results for infinitely long systems with perfect helical symmetry can be applied to finite-length systems. Meanwhile, as demonstrated above, this boundary-integral method with reduced dimension can be applied to studying other more complicated fluid effects and can be extended to non-uniform structures. More results with application to these systems will be reported in future. }

\begin{acknowledgments} 
This work was supported by National Science Foundation Grant No. CBET-0854108.
\end{acknowledgments}

\appendix
\section{Integral kernel in the body-centerline coordinates}
In Cartesian coordinates, a point on a surface $D$ with helical symmetry can be written as
\begin{equation}
\mathbf{x} (\varphi, \alpha)=R \left(\begin{array}{c} 
\cos\varphi\\
\sin\varphi\\
\varphi/\tan(\theta)
\end{array}\right)
+\rho \mathcal{R}^3(\varphi) \mathcal{R}^1(-\theta)\left(\begin{array}{c} 
\cos\alpha\\
\sin\alpha\\
0
\end{array}\right),
\end{equation}
where $R$ and $\varphi$ are the radius and the phase angle of the helix, $\rho$, $\alpha$ are the radius and the polar angle of the surface $D$ intercepted by a plane (see Fig.~\ref{fig:2}). According to the definition of the Stokeslet (Eq.~\ref{greens}) and using the convention that $\varphi_C=0$, the integral kernel for modified Stokeslets $\tilde{\mathcal{H}}_{ji}(\mathbf{q}(\alpha'), \mathbf{q}(\alpha))$ (expressed in Eq.~(\ref{bi-bc02_2})) can be expressed as
\begin{equation}
\mathcal{I}=\mathcal{P}(\theta, \varphi')\cdot \mathcal{G}(\mathbf{x}(\varphi', \alpha'), \mathbf{x}(0, \alpha)) \cdot\mathcal{P}^{-1}(\theta, 0)=\frac{\tilde{\mathcal{R}}^z(\varphi')}{d}+\frac{\mathbf{Z}_1 \mathbf{Z}_2}{d^3} . \label{s-kernel}
\end{equation}
Here, $\tilde{\mathcal{R}}^z$ is a rotation operator defined in the body-centerline coordinates  $\mathbf{q}$:
\begin{equation}
\tilde{\mathcal{R}}^z(\varphi')=\mathcal{R}^1(\theta)\cdot \mathcal{R}^3(\varphi')\cdot\mathcal{R}^1(-\theta),
\end{equation}
and vectors $\mathbf{Z}_1$, $\mathbf{Z}_2$ are projections of vector ($\mathbf{x}(\varphi', \alpha')-\mathbf{x}(0, \alpha)$) in the body-centerline coordinates. More specifically,
\begin{eqnarray}
\mathbf{Z}_1=&&\mathcal{P}(\theta, \varphi')\cdot \left(\mathbf{x}(\varphi', \alpha')-\mathbf{x}(0, \alpha)\right)\\
=&&\left(\begin{array}{l}
\rho(\alpha')(
\cos\varphi' \cos\alpha'-\sin\varphi'\cos\theta\sin\alpha')-\rho (\alpha)\cos\alpha + R(\cos\varphi'-1)\\
\rho(\alpha')\left(\cos\theta\sin\varphi'\cos\alpha'+\frac{(\cos\varphi'-1)\cos2\theta+(\cos\varphi'+1)\sin\alpha'}{2}\right)-\rho(\alpha)\sin\alpha\\
\quad +R(\sin\varphi'-\varphi')\cos\theta\\
\rho(\alpha')\left(\sin\varphi'\sin\theta\cos\alpha'+\frac{(\cos\varphi'-1)\sin2\theta\sin\alpha'}{2}\right)+R\sin\theta\sin\varphi'\\
\quad +\varphi'\Gamma\cos^2\theta /\pi
\end{array}\right),
\end{eqnarray}
and
\begin{eqnarray}
\mathbf{Z}_2=&&\mathcal{P}(\theta, 0)\cdot \left(\mathbf{x}(\varphi', \alpha')-\mathbf{x}(0, \alpha)\right)\\
=&&\left(\begin{array}{l}
\rho (\alpha')\cos\alpha'-\rho(\alpha)(
\cos\varphi' \cos\alpha+\sin\varphi'\cos\theta\sin\alpha) + R(1-\cos\varphi')\\
\rho(\alpha')\sin\alpha'+\rho(\alpha)\left(\cos\theta\sin\varphi'\cos\alpha-\frac{(\cos\varphi'-1)\cos2\theta+(\cos\varphi'+1)\sin\alpha}{2}\right)\\
\quad +R(\sin\varphi'-\varphi')\cos\theta\\
\rho(\alpha)\left(\sin\varphi'\sin\theta\cos\alpha-\frac{(\cos\varphi'-1)\sin2\theta\sin\alpha}{2}\right)+R\sin\theta\sin\varphi'\\
\quad +\varphi'\Gamma\cos^2\theta /\pi
\end{array}\right).
\end{eqnarray}

\section{Singularity reduction}

\begin{figure}
\begin{center}
\includegraphics[width=0.6\textwidth]{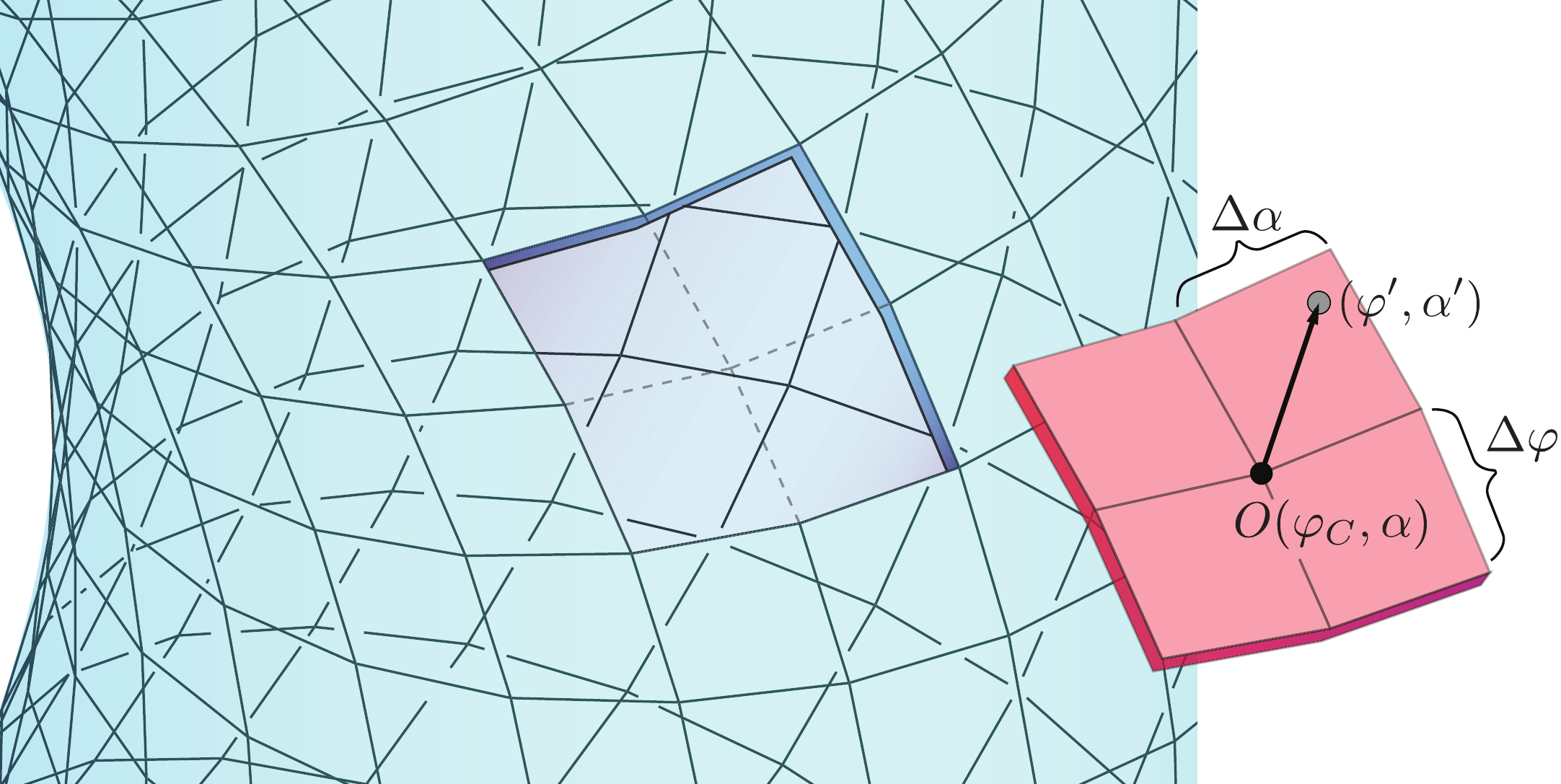}
\caption{(Color online.) Singularity reduction.  Boundary integral on a meshed surface near the singular origin $O$ (shown as the cropped section) is valued analytically. The integral on the rest of the surface is performed numerically using the trapezoidal rule. 
\label{fig:7}}
\end{center}
\end{figure}
In order to avoid the numerical divergence of the singular kernel $\mathcal{G}(\mathbf{x}, \mathbf{x}_C)$, we evaluate the integral separately for the modified Stokeslet $\tilde{\mathcal{H}}$ when $d=|\mathbf{x}-\mathbf{x}_C|$ is below grid size.  As shown in Fig.~\ref{fig:7}, a small patch of surface (defined by the phase angel $\varphi  \in [\varphi_C-\Delta\varphi, \varphi_C+\Delta\varphi]$ and polar angle $\alpha \in [\alpha-\Delta \alpha, \alpha+\Delta \alpha]$) is cropped out. The integral over the rest of the surface is performed numerically, using the trapezoidal rule. We evaluate the integral over that small patch in an analytical form, to the order of accuracy that is no worse than $\mathcal{O}(\Delta\varphi^3, \Delta \alpha^3)$. 

To formulate an infinitesimal expansion of such integral, we consider an arbitrary point on the patch $\mathbf{x}(\delta \varphi, \alpha+\delta \alpha)$, with a small displacement from its center $\mathbf{x}(0, \alpha)$. The displacement can be written as 
\begin{eqnarray}
&&\mathbf{x} (\delta \varphi, \alpha+\delta \alpha)-\mathbf{x} (0, \alpha)=\\
&&R \left[(\mathcal{R}^3(\delta \varphi)-\mathbb{I})\cdot\left(\begin{array}{c} 
1\\
0\\
0
\end{array}\right)
+\frac{\delta\varphi}{\tan\theta}\left(\begin{array}{c} 
0\\
0\\
1
\end{array}\right)
\right]\\
&&+a \left[\mathcal{R}^3(\delta\varphi) \mathcal{R}^1(-\theta)\mathcal{R}^3(\delta \alpha)- \mathcal{R}^1(-\theta)\right]\left(\begin{array}{c} 
\cos\alpha\\
\sin\alpha\\
0
\end{array}\right),
\end{eqnarray}
where the second term on the right-hand side is due to surface curvature, and can be expanded in infinitesimal form as
\begin{eqnarray}
&&\mathcal{R}^3(\delta\varphi) \mathcal{R}^1(-\theta)\mathcal{R}^3(\delta \alpha)- \mathcal{R}^1(-\theta)\\
&&=\delta\varphi \mathbf{\epsilon} \cdot \mathcal{R}^1(-\theta)+\delta \alpha \mathcal{R}^1(-\theta)\cdot \mathbf{\epsilon} + \mathcal{O}([\delta\varphi, \delta\alpha]^2),
\end{eqnarray}
where tensor $\epsilon$ is defined as 
\begin{eqnarray}
\epsilon=\left(\begin{array}{ccc}
0 & -1 & 0\\
1 & 0 & 0\\
0 & 0 & 0
\end{array}\right)
\end{eqnarray}
To the lowest order of infinitesimal expansion, vectors $\mathbf{Z}_1$ and $\mathbf{Z}_2$ in Eq.~(\ref{s-kernel}) become 
\begin{eqnarray}
\mathbf{Z}_1 = \mathbf{Z}_2&=&\mathcal{R}^1(\theta)\cdot \left[\vec{x} (\delta \varphi, \alpha+\delta \alpha)-\vec{x} (0, \alpha)\right]\\
&=& a\delta \varphi \mathbf{e}_\varphi +a\delta \alpha \mathbf{e}_\alpha+ \mathcal{O}([\delta\varphi, \delta \alpha]^2),
\end{eqnarray}
where vectors 
\begin{equation}\textrm{$\mathbf{e}_\varphi=${\scriptsize$\left(\begin{array}{c} 
-\cos\theta \sin \alpha\\
\cos \alpha \cos \theta\\
\cos\alpha \sin \theta + \frac{R}{a}\csc\theta
\end{array}\right)$}, \mbox{and}
$\mathbf{e}_\alpha=${\scriptsize$\left(\begin{array}{c} 
-\sin \alpha\\
\cos \alpha\\
0
\end{array}\right)$}}.
\end{equation}
By introducing 
\begin{eqnarray}
\left(\begin{array}{c}
X_1\\
X_2
\end{array}
\right)
=\left(\begin{array}{cc}
1 & b_1\\
0 & b_2
\end{array}
\right)
\left(\begin{array}{c}
\delta \alpha\\
\delta \varphi
\end{array}
\right),\label{vtrsf}
\end{eqnarray}
with $b_1=\cos\theta$ and $b_2= \left|\sin\theta \cos \alpha+\frac{R}{a} \csc\theta  \right|$, the rescaled distance $|\mathbf{Z}_1|/a$ (or $|\mathbf{Z}_2|/a$), can be expressed quadratically as 
\begin{eqnarray}
&&\xi(\delta \varphi, \delta \alpha) = \frac{1}{a}\left|\mathbf{x} (\delta \varphi, \alpha+\delta \alpha)-\mathbf{x} (0, \alpha)\right|=\sqrt{{X_1}^2+{X_2}^2}.
\end{eqnarray}

\noindent This quadratic form facilitates the analytic integration of the singular kernel,
\begin{eqnarray}
\mathcal{I}=\frac{1}{a}\left(\frac{\mathbb{I}}{\xi}+\frac{(\delta \varphi \mathbf{e}_\varphi +\delta \alpha \mathbf{e}_\alpha)(\delta \varphi \mathbf{e}_\varphi +\delta \alpha \mathbf{e}_\alpha)}{\xi^3}\right)+\mathcal{O}([\delta\varphi, \delta\alpha]^2),
\end{eqnarray}
by transforming the integral domain to polar coordinates. The integral can thus be obtained as
\begin{eqnarray}
&&\int_{-\Delta \varphi}^{\Delta\varphi}\textrm{d}\delta\varphi\int_{-\Delta \alpha}^{\Delta \alpha}\textrm{d}\delta\alpha ~\mathcal{I}\cdot \tilde{\mathbf{f}}^J(\alpha)\\
&&=\frac{1}{a}\left[B_0 \mathbb{I}+B_{\varphi\varphi}\mathbf{e}_\varphi\mathbf{e}_\varphi+B_{\alpha\alpha}\mathbf{e}_\alpha\mathbf{e}_\alpha+B_{\varphi\alpha}(\mathbf{e}_\varphi\mathbf{e}_\alpha+\mathbf{e}_\alpha\mathbf{e}_\varphi)\right]\cdot \tilde{\mathbf{f}}^J(\alpha) \nonumber \\
&&+\mathcal{O}([\Delta \varphi, \Delta \alpha]^3). \nonumber
\end{eqnarray}
The error of order $\mathcal{O}([\Delta\varphi, \Delta \alpha]^2)$ in the above formulation vanishes due to geometrical reasons. The factors $B_0$, $B_{\varphi\varphi}$, $B_{\alpha\alpha}$ and $B_{\varphi \alpha}$ are obtained analytically as the following:
\begin{eqnarray}
B_0&\equiv&\int_{-\Delta \alpha}^{\Delta \alpha} \textrm{d}\delta \alpha \int_{-\Delta \varphi}^{\Delta \varphi} \textrm{d}\delta \varphi\frac{1}{\xi}
=\frac{1}{b_2}\iint_{\Delta S}\textrm{d}X \textrm{d}Y\frac{1}{\sqrt{X^2+Y^2}}\\
&=&\frac{1}{ b_2}\left\{D_1 \ln \left[\frac{\left(1+\sin(\gamma_2-\gamma_0)\right)\left(1-\sin(\gamma_1-\gamma_0)\right)}{\left(1-\sin(\gamma_2-\gamma_0)\right)\left(1+\sin(\gamma_1-\gamma_0)\right)}\right] \right. \nonumber\\ 
&&\left.+D_2 \ln \left[\frac{\left(1-\cos\gamma_3\right)\left(1+\cos\gamma_2\right)}{\left(1+\cos\gamma_3\right)\left(1-\cos\gamma_2\right)}\right]\right\},\nonumber
\end{eqnarray}
\begin{eqnarray}
B_{\varphi\varphi} &\equiv &\int_{-\delta \alpha}^{\delta \alpha} \textrm{d}\Delta \alpha \int_{-\delta \varphi}^{\delta \varphi} \textrm{d}\Delta \varphi\frac{\Delta \varphi^2 }{\xi^3}
=\frac{1}{b_2^3}\iint \textrm{d}X \textrm{d}Y\frac{Y^2}{{(X^2+Y^2)}^{3/2}}\\
&=&\frac{1}{b_2^3} \biggr\{ 2D_1 \left[\sin(\gamma_1+\gamma_0)-\sin(\gamma_2+\gamma_0)\right]  \nonumber\\
&&+D_1 \cos^2\gamma_0 \ln\left[\frac{\left(1+\sin(\gamma_2-\gamma_0)\right)\left(1-\sin(\gamma_1-\gamma_0)\right)}{\left(1-\sin(\gamma_2-\gamma_0)\right)\left(1+\sin(\gamma_1-\gamma_0)\right)}\right] \nonumber\\
&&-2D_2 \left(\cos\gamma_3-\cos\gamma_2\right) \biggr\}, \nonumber
\end{eqnarray}
\begin{eqnarray}
B_{\alpha\alpha}&\equiv&\int_{-\delta \alpha}^{\delta \alpha} \textrm{d}\Delta \alpha \int_{-\delta \varphi}^{\delta \varphi} \textrm{d}\Delta \varphi\frac{\Delta \alpha^2 }{\xi^3}
=\frac{1}{ b_2}\iint \textrm{d}X \textrm{d}Y\frac{\left(X-\frac{b_1}{b_2}Y\right)^2}{{(X^2+Y^2)}^{3/2}}\\
&=&\frac{1}{b_2} \biggr\{ 2D_1 \sec^2\gamma_0\left[ \sin\left(\gamma_2-\gamma_0\right)-\sin\left(\gamma_1-\gamma_0\right)\right] \nonumber\\
&&+D_2 \left[2\left(1-\frac{b_1^2}{b_2^2}\right) (\cos\gamma_3 - \cos\gamma_2) - \frac{4b_1}{b_2} (\sin \gamma_3 - \sin \gamma_2) \right] \nonumber\\
&&-D_2 \ln \left[\frac{(1+\cos\gamma_3)(1-\cos\gamma_2)}{(1-\cos\gamma_3)(1+\cos\gamma_2)}\right] \biggr\}, \nonumber
\end{eqnarray}
and
\begin{eqnarray}
B_{\varphi\alpha}&\equiv&\int_{-\delta \alpha}^{\delta \alpha} \textrm{d}\Delta \alpha \int_{-\delta \varphi}^{\delta \varphi} \textrm{d}\Delta \varphi\frac{\Delta \alpha\Delta \varphi}{\xi^3}
=\frac{1}{ b_2^2}\iint \textrm{d}X \textrm{d}Y\frac{\left(X-\frac{b_1}{b_2}Y\right)Y}{{(X^2+Y^2)}^{3/2}} \\
&=&\frac{1}{ b_2^2} \biggr\{ 2 D_1\sec \gamma_0 \left(\cos\gamma_1-\cos\gamma_2\right) \nonumber\\
&&+2 D_2 \sec\gamma_0 \left[\sin (\gamma_3 - \gamma_0) - \sin(\gamma_2-\gamma_0)\right] 
\biggr\}, \nonumber
\end{eqnarray}
where 
\begin{eqnarray}
D_1&=&\cos\gamma_0\Delta \alpha, \quad D_2=b_2 \Delta \varphi, \\
\gamma_0 &=& \tan^{-1}\left(-\frac{b_1}{b_2}\right),\\
\gamma_1 &=& \tan^{-1}\left(-b_2\Delta \varphi/(\Delta \alpha - b_1\Delta \varphi)\right),\\
\gamma_2 &=& \tan^{-1}\left(b_2\Delta \varphi/(\Delta \alpha + b_1\Delta \varphi)\right), \mbox{ and}\\
\gamma_3 &=& \pi +\gamma_1.
\end{eqnarray}
In this way, the singularity in the boundary integral is 
avoided. Even though the integral kernel $\mathcal{I}$ is singular, its integral near the origin converges. The 
resultant integral, the term, $\mathcal{E}_m$, that compensates for the singularity removal (in Eq.~\ref{bi-bc02_3}), is linearly proportional to the factors ($D_1$, $D_2$), and thus linearly proportional to the mesh size ($\Delta \varphi$, $\Delta \alpha$). 

\section{Convergence analysis}

\begin{figure}
\begin{center}
\includegraphics[width=0.75\textwidth]{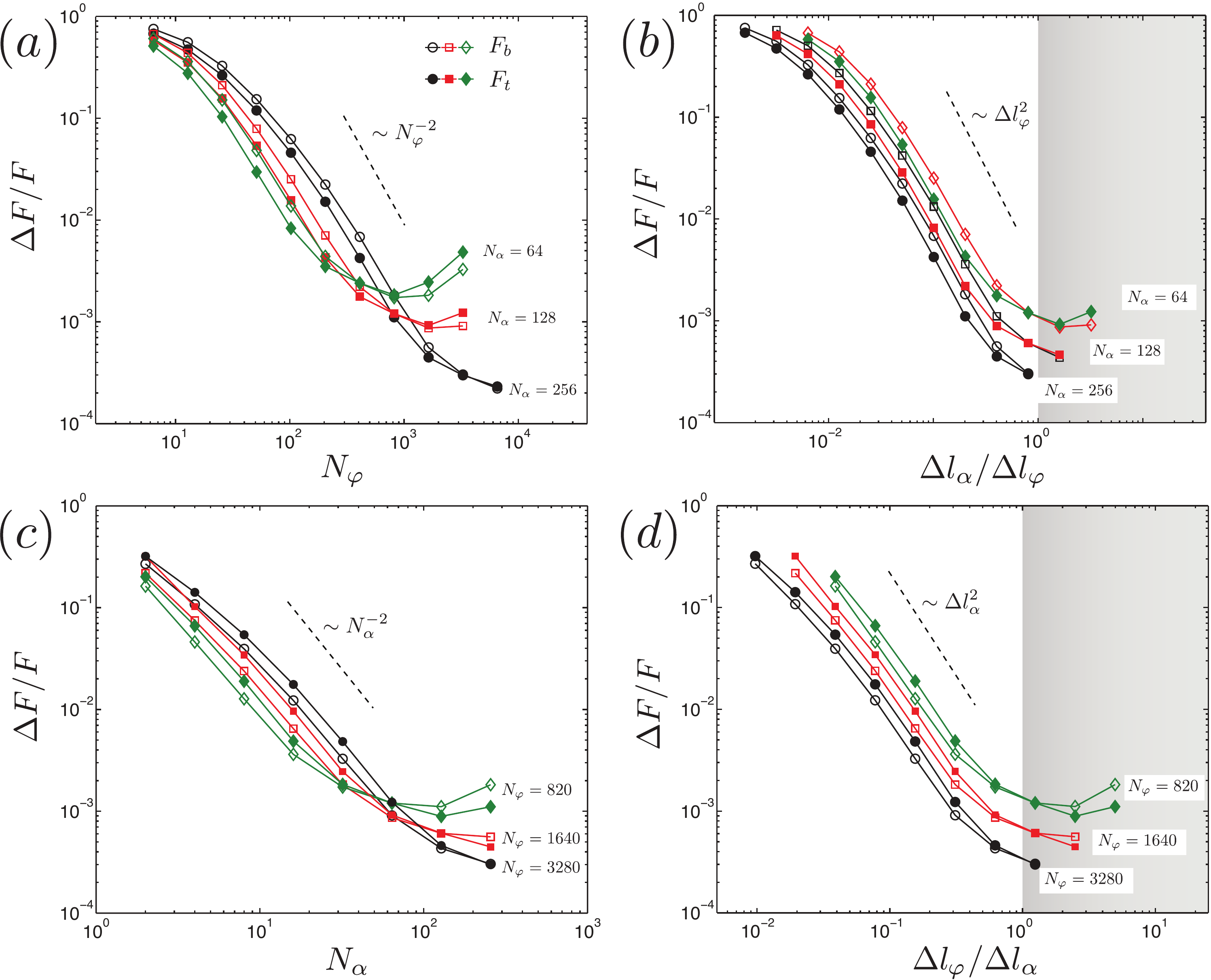}
\caption{(Color online.) Numerical convergence of the hydrodynamic force per unit arc-length $\tilde{\mathbf{F}}=(0, F_b, F_t)$. The helical geometry is given by the ratio of the radius of the cross-section to its arc length per pitch $a/\Gamma=0.01$, pitch angle $\theta=\pi/4$, and number of pitches $\kappa = 40$. (a) Relative error of the force components $\Delta F/F$ as a function of the number of grids $N_\varphi$ with fixed $N_\alpha$. (b) Re-plot of data in (a) against $\Delta l_\alpha/\Delta l_\varphi$, the ratio between the physical length scales regarding grids $\Delta\alpha$ and $\Delta\varphi$. The error starts to increase when $N_\varphi$ is sufficiently large or when $\Delta l_\varphi$ becomes less than $\Delta l_\alpha$, as shadowed in gray. (c) Relative error of the force components  as a function of the number of grids $N_\alpha$. (d) Re-plot of data in (c) against $\Delta l_\varphi/\Delta l_\alpha$. Similar to (a) and (b), the solution diverges when $N_\alpha$ is sufficiently large or when $\Delta l_\alpha<\Delta l_\varphi$, as shadowed in gray. The dashed lines in each figure indicate second-order convergence. \label{fig:8}}
\end{center}
\end{figure}
To further benchmark this modified boundary element method, we study the 
numerical convergence of the associated solutions. 
One commonly-computed feature of helical propulsion is the hydrodynamic force per unit arc-length, defined as
\begin{equation}
\tilde{\mathbf{F}}=\sum_{l=1}^{N_\alpha} \Delta \alpha \tilde{\mathbf{f}}^J(\alpha_l),
\end{equation}
where ($\tilde{F}_1$, $\tilde{F}_2$, $\tilde{F}_3$)=($F_n$, $F_b$, $F_t$), and $F_t$, $F_n$, $F_b$ are the components tangential, normal, and binormal to the body-centerline, respectively. \blbr{To check the convergence, we compute the components of the force using successively better spatial resolution $(N_\varphi, N_\alpha)$. The errors $\Delta F$ of these components are defined by subtracting the forces from the forces computed at the highest resolution.}
Figure~\ref{fig:8} shows a typical result of such analysis for a tethered helix. Here, $\kappa=40$, $\theta=\pi/4$, and $a/\Gamma=0.01$. It should be note that $F_n=0$ due to geometrical reasons, and is thus independent of the spatial resolution. For fixed resolution along the circumference ($N_\alpha$), the numerical errors ($\Delta F_t$, $\Delta F_b$) decrease as $N_\varphi$ increases, approaching second-order convergence ($\Delta F \sim N_\varphi^{-2}$).
However, when $N_\varphi$ 
reaches a threshold, the numerical errors start to increase. This threshold for divergence is given by the condition that the mesh along $\varphi$ is so dense that the associated length scale of the mesh size, $\Delta l_\varphi \approx N_\lambda \Gamma/N_\varphi$, is smaller than that along the circumference, $\Delta l_\alpha = 2\pi a/N_\alpha$. It is 
plausible that this feature of convergence is due to the fact that we dissect the original boundary integral equation (Eq.~(\ref{bi-cartesian})) in two steps (Eq.~(\ref{bi-bc01_3}) and Eq.~(\ref{bi-bc02_3})). Note that the force densities at different locations but along the helical contour (see dashed curve in Fig.~\ref{fig:1}) are assumed to be identical, the correlations among these sites are treated 
as ``self-interaction". When the grids along $\varphi$ are too dense, the helical symmetry is ``exaggerated", and the effective self-interaction,
$\tilde{\mathcal{H}}(n, n)$, obtained from Eq.~(\ref{bi-bc02_3}) also diverges with a $\mathcal{O}(\log(1/N_\varphi))$ dependence. 
\begin{figure}
\begin{center}
\includegraphics[width=0.75\textwidth]{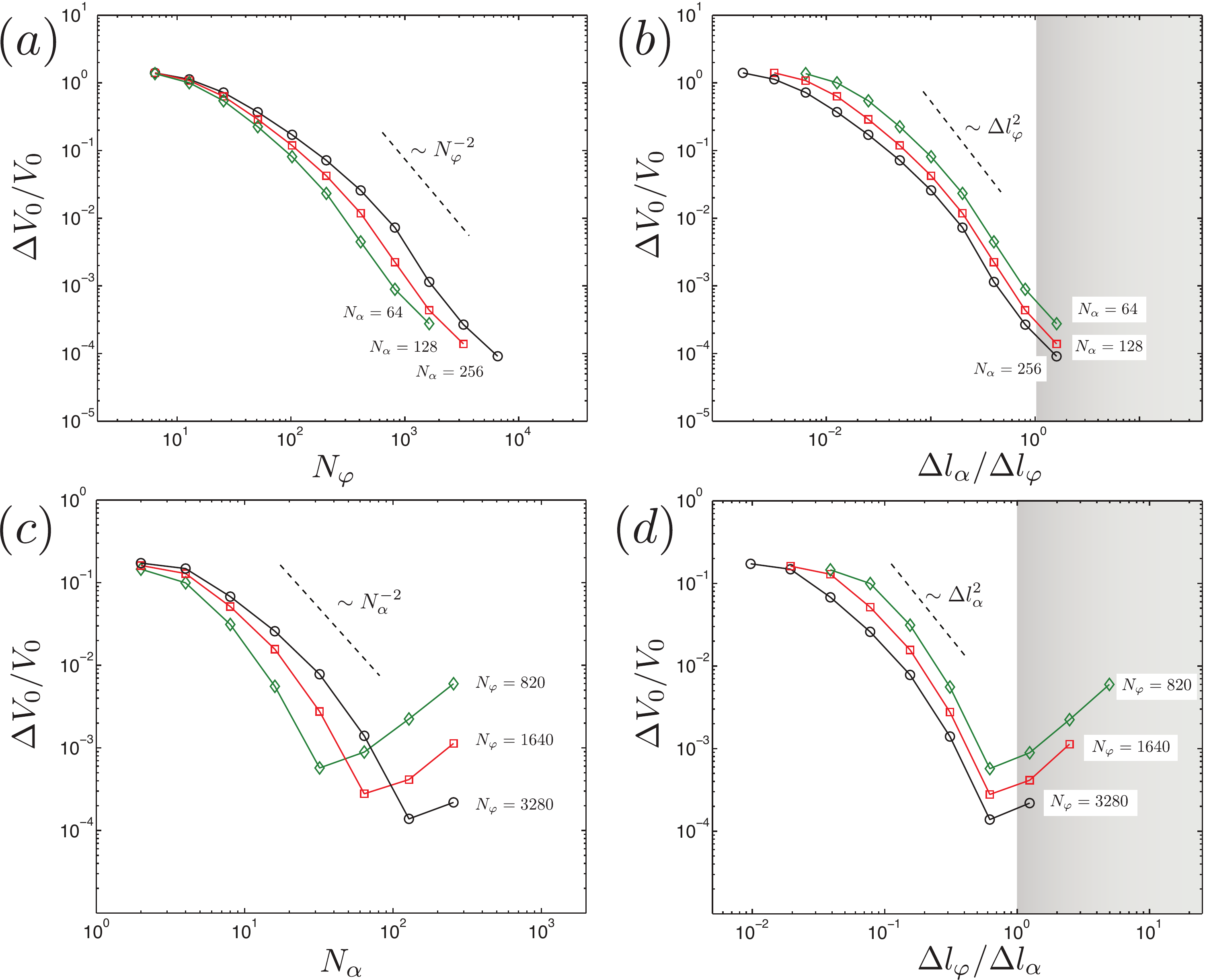}
\caption{(Color online.) Motility of a force-free helix $V_0$ and its numerical convergence. Helical filament with the same geometry as shown in Fig.~\ref{fig:8} are used. (a) Relative error of the free-swimming speed $\Delta V_0/V_0$ as a function of the number of grids $N_\varphi$. (b) Re-plot of data in (a) against $\Delta l_\alpha/\Delta l_\varphi$. (c) Relative error as a function of the number of grids $N_\alpha$. (d) Re-plot of data in (c) against $\Delta l_\varphi/\Delta l_\alpha$. Similar dependencies as shown in convergence of force components (Fig.~\ref{fig:8}) are observed: the simulation result becomes divergent if grids along $\varphi$ (or $\alpha$) are too dense, as characterized by the ratio $\Delta l_\alpha/\Delta l_\varphi$. The dashed lines in each figure indicate second-order convergence.
\label{fig:9}}
\end{center}
\end{figure}

On the other hand, if we fix the number of grid points, $N_\varphi$, $\Delta F$ also 
exhibits second order convergence until $N_\alpha$ is below a critical value. Again, this critical value of $N_\alpha$ is determined by the criterion $\Delta l_\varphi \sim \Delta l_\alpha$. 

The numerical error in the free-swimming speed, $V_0(N_\varphi, N_\alpha)$, for a helix of fixed geometry ($a/\Gamma$=0.026, $\theta$=$\pi/4$, $\kappa$=40) is shown in Fig.~\ref{fig:9}. Similar to the hydrodynamic force on a tethered helix, $V_0$ converges with second-order accuracy as the number of mesh points grows along either of the two directions ($\varphi$ or $\alpha$). 
However, as was observed with the force, $\Delta f$, the numerical error in $\Delta V_0$ also starts to increase, when $\Delta l_\varphi \sim \Delta l_\alpha$ (see Fig.~\ref{fig:8}(b) and (d)).
\begin{figure}
\begin{center}
\includegraphics[width=0.75\textwidth]{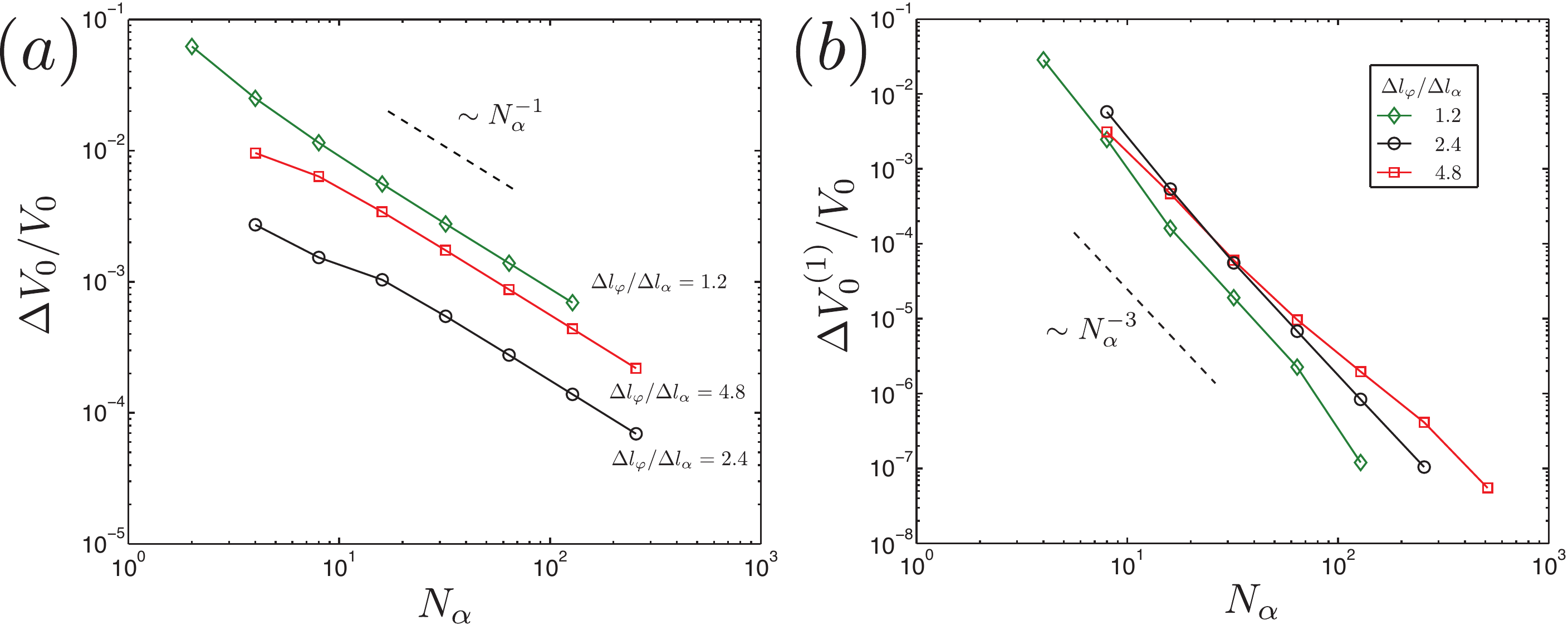}
\caption{(Color online.) (a) Numerical convergence of the free-swimming $V_0$ of a helix (with the same geometry as shown in Fig.~\ref{fig:8} and Fig.~\ref{fig:9}) with fixed ratio $\Delta l_\varphi/\Delta l_\alpha$. The dashed line shows a first order convergence. (b) Convergence of the free-swimming with first-order error subtracted using Eq.~(\ref{eq:v3rd}). The dashed line indicates a third order convergence.
\label{fig:10}}
\end{center}
\end{figure}

Since the divergence in the numerical error is governed by the ratio $\Delta l_\alpha /\Delta l_\varphi$, one way to ensure a robust convergence is to increase both $N_\varphi$ and $N_\alpha$ simultaneously so that $\Delta l_\alpha /\Delta l_\varphi$ is fixed. As shown in Fig~\ref{fig:10}(a), for fixed $\Delta l_\alpha /\Delta l_\varphi$, the numerical error $\Delta V_0$ shows a robust convergence, regardless on the value of $\Delta l_\alpha /\Delta l_\varphi$ that we choose. As a tradeoff for this robustness, the numerical convergence becomes first order. \blbr{We suspect that 
this lower-order convergence arises because our singularity reduction technique does not completely respect the helical symmetry, which in turn leads to a  modified Stokeslet $\tilde{\mathcal{H}}$  (Eq.~(\ref{bi-bc02_3})) that is less accurate than second order.} However, such numerical convergence can be improved by a Richardson extrapolation of the above $V_0(N_\varphi, N_\alpha)$ as
\begin{equation}
V_0^{(1)}(N_\varphi, N_\alpha) = 2V_0(N_\varphi, N_\alpha)-V_0\left([N_\varphi/2], [N_\alpha/2]\right). \label{eq:v3rd}
\end{equation}
The resulting swimming speed, $V_0^{(1)}$, 
now exhibits third-order convergence (Fig.~\ref{fig:10}(b)) with marginal additional computational cost. 
\blbr{While not shown here, the convergence of computed force at fixed ratio $\Delta l_\alpha /\Delta l_\varphi$ is similar to the above case of $V_0$.}


\begin{thebibliography}{10}

\bibitem{Pozrikidis:1992td}
C.~Pozrikidis,
\newblock {\em Boundary Integral and Singularity Methods for Linearized Viscous
  Flow} (Cambridge University Press, Cambridge, England, 1992).

\bibitem{Lauga:RPP2009}
E.~Lauga and T.~R. Powers,
\newblock ``The hydrodynamics of swimming microorganisms,"
\newblock Rep. Prog. Phys. {\bf 72}, 096601 (2009).

\bibitem{Phan-Thien:1987}
N.~Phan-Thien, T.~Tran-Cong, and M.~Ramia,
\newblock ``A boundary-element analysis of flagellar propulsion,"
\newblock J. Fluid Mech. {\bf 184}, 533 (1987).

\bibitem{Power:1987}
H.~Power and G.~Miranda,
\newblock ``Second kind integral equation formulation of Stokes' flows past a particle of arbitrary shape,"
\newblock SIAM J. Appl. Math. {\bf 47}, 689 (1987).

\bibitem{Gonzalez:2009}
O.~Gonzalez,
\newblock ``On stable, complete, and singularity-free boundary integral formulations of exterior Stokes flow,"
\newblock SIAM J. Appl. Math. {\bf 69}, 933 (2009).

\bibitem{Keaveny:JCP2011}
E.~E. Keaveny and M.~J. Shelley,
\newblock ``Applying a second-kind boundary integral equation for surface tractions in Stokes flow,"
\newblock J. Comp. Phys. {\bf 230}, 2141 (2011).

\bibitem{Cortez:2001}
R.~Cortez,
\newblock ``The method of regularized Stokeslets,"
\newblock SIAM J. Sci. Comput. {\bf 23}, 1204 (2001).

\bibitem{Cortez:PF2005}
R.~Cortez, L.~Fauci, and A.~Medovikov,
\newblock ``The method of regularized Stokeslets in three dimensions: Analysis, validation, and application to helical swimming,"
\newblock Phys. Fluids {\bf 17}, 031504 (2005).

\bibitem{Smith:2009kh}
D.~J. Smith,
\newblock ``A boundary element regularized Stokeslet method applied to cilia- and flagella-driven flow,"
\newblock Proc. R. Soc. A {\bf 465}, 3605 (2009).

\bibitem{Bouzarth:2011}
E.~L. Bouzarth and M.~L. Minion,
\newblock ``Modeling slender bodies with the method of regularized Stokeslets,"
\newblock J. Comput. Phys. {\bf 230}, 3929 (2011).

\bibitem{Berg:1973td}
H.~Berg,
\newblock ``Bacteria swim by rotating their flagellar filaments,"
\newblock Nature {\bf 245}, 380 (1973).

\bibitem{Charon:1984}
N.~Charon, G.~Daughtry, R.~Mccuskey, and G.~Franz,
\newblock ``Microcinematographic analysis of tethered Leptospira illini,"
\newblock J. Bacteriol. {\bf 160}, 1067 (1984).

\bibitem{Horridge:1969vt}
G.~A. Horridge and G.~A. Tamm,
\newblock ``Critical point drying for scanning electron microscopic study of ciliary motion,"
\newblock Science {\bf 163}, 817 (1969).

\bibitem{Purcell:1977}
E.~M. Purcell,
\newblock ``Life at low Reynolds-number,"
\newblock Am. J. Phys. {\bf 45}, 3 (1977).

\bibitem{Thaokar:2007}
R.~M. Thaokar, H.~Schiessel, and I.~M. Kulic,
\newblock ``Hydrodynamics of a rotating torus,"
\newblock Eur. Phys. J. B {\bf 60}, 325 (2007).

\bibitem{Spagnolie:2010}
S.~E. Spagnolie and E.~Lauga,
\newblock ``Jet propulsion without inertia,"
\newblock Phys. Fluids {\bf 22}, 081902 (2010).

\bibitem{Lighthill:1976}
J.~Lighthill,
\newblock ``Flagellar Hydrodynamics: The John von Neumann Lecture, 1975,"
\newblock SIAM Rev. {\bf 18}, 161 (1976).

\bibitem{lighthill1996b}
J.~Lighthill,
\newblock ``Reinterpreting the basic theorem of flagellar hydrodynamics,"
\newblock J. Eng. Math. {\bf {\bf 30}}, 25 (1996).

\bibitem{Childress:1989}
S.~Childress, M.~Landman, and H.~Strauss,
\newblock ``Steady motion with helical symmetry at large reynolds number,"
\newblock in {\em Proc. IUTAM Symp. on Topological Fluid dynamics}, edited by
  H.~K. Moffatt and A.~Tsinober, 216-224 (Cambridge University Press, Cambridge, England, 1989).

\bibitem{Zabielski:1998}
L.~Zabielski and A.~Mestel,
\newblock ``Steady flow in a helically symmetric pipe,"
\newblock J. Fluid Mech. {\bf 370}, 297 (1998).

\bibitem{Delbende:2011}
I.~Delbende, M.~Rossi, and O.~Daube,
\newblock ``DNS of flows with helical symmetry,"
\newblock Theor. Comput. Fluid Dyn. {\bf 26}, 141 (2012).

\bibitem{kimbook}
S.~Kim and J.~S. Karrila,
\newblock {\em Microhydrodynamics: {P}rinciples and {S}elected {A}pplications.},
\newblock (Butterworth-Heinemann, Newton, MA, 1991).

\bibitem{Powers2010}
T.~R. Powers,
\newblock ``Dynamics of filaments and membranes in a viscous fluid,"
\newblock Rev. Mod. Phys. {\bf \textbf{82}}, 1607 (2010).

\bibitem{Liu:PNAS2011}
B.~Liu, T.~R. Powers, and K.~S. Breuer,
\newblock ``Force-free swimming of a model helical flagellum in viscoelastic fluids,"
\newblock Proc. Natl. Acad. Sci. U. S. A. {\bf 108}, 19516 (2011).

\bibitem{Gray:1955}
J.~Gray and G.~J. Hancock,
\newblock ``The propulsion of sea-urchin spermatozoa,"
\newblock J. Exp. Biol. {\bf 32}, 802 (1955).

\bibitem{Childress1981}
S.~Childress,
\newblock {\em Mechanics of swimming and flying},
\newblock (Cambridge University Press, Cambridge, England, 1981).

\bibitem{JOHNSON:1980p98}
R.~Johnson,
\newblock ``An improved slender-body theory for Stokes flow,"
\newblock J. Fluid. Mech. {\bf 99}, 411 (1980).

\bibitem{Rodenborn:2013}
B.~Rodenborn, C.-H.~Chen, H.~L.~Swinney, B.~Liu, and H.~P.~Zhang,
\newblock ``Propulsion of microorganisms by a helical flagellum,"
\newblock Proc. Natl. Acad. Sci. U.~S.~A. {\bf 110}, E338 (2013).

\bibitem{keller_rubinow1976}
J.~B. Keller and S.~I. Rubinow,
\newblock ``Slender body theory for viscous flow,"
\newblock J. Fluid Mech. {\bf {\bf 75}}, 705 (1976).

\end{thebibliography}
\end{document}